\documentclass{aa}
\usepackage{graphicx}

\usepackage{natbib}

\usepackage[german,english]{babel}
\newcommand{\gppr}{\stackrel{>}{\scriptstyle \sim}}
\newcommand{\gappr}{\raisebox{-0.4ex}{$\gppr$}}
\newcommand{\lppr}{\stackrel{<}{\scriptstyle \sim}}
\newcommand{\lappr}{\raisebox{-0.4ex}{$\lppr$}}

\newcommand{\Jcmb}{\mbox{$\dot{J}_{\mathrm{CMB}}$}}
\newcommand{\Jrmb}{\mbox{$\dot{J}_{\mathrm{RMB}}$}}
\newcommand{\Ocri}{\mbox{$\omega_\mathrm{crit}$}}
\newcommand{\Psd}{\mbox{$P_\mathrm{sd}$}}
\newcommand{\tsd}{\mbox{$t_\mathrm{sd}$}}
\newcommand{\Porb}{\mbox{$P_\mathrm{orb}$}}

\newcommand{\Mwd}{\mbox{$M_\mathrm{1}$}}
\newcommand{\Msec}{\mbox{$M_\mathrm{sec}$}}
\newcommand{\Rwd}{\mbox{$R_\mathrm{1}$}}
\newcommand{\Rsec}{\mbox{$R_\mathrm{sec}$}}
\newcommand{\Twd}{\mbox{$T_\mathrm{eff}$}}
\newcommand{\Ha}{\mbox{${\mathrm H\alpha}$}}
\newcommand{\Msun}{\mbox{$M_{\odot}$}}
\newcommand{\Rsun}{\mbox{$R_{\odot}$}}
\newcommand{\Teff}{\mbox{$T_\mathrm{eff}$}}
\newcommand{\La}{\mbox{${\rm Ly\alpha}$}}
\newcommand{\ecsa}{\mbox{$\rm erg\;cm^{-2}s^{-1}\mbox{\AA}^{-1}$}}

\begin{document}

\title{The age, life expectancy, and space density of \\ Post Common
Envelope Binaries}

\author{M.R. Schreiber \inst{1} \and B.T. G\"ansicke \inst{2}}

\offprints{M.R. Schreiber,
\email{mschrei@astro.u-strasbg.fr}}

\institute{Universit{\'e} Louis Pasteur, Observatoire de Strasbourg, 
11 rue de l'Universit{\'e}, 67000 Strasbourg, France 
\and 
Department of Physics and Astronomy, University of Southampton,
Highfield, Southampton SO17 1BJ, UK}

\date{Received \underline{\hskip2cm} ; accepted \underline{\hskip2cm} }

\titlerunning{PCEBs}

\authorrunning{Schreiber\,\&\,G\"ansicke}

\abstract{
We present a sample of 30 well observed Post Common Envelope Binaries
(PCEBs). Deriving the cooling age of the white dwarfs, we show that
the PCEB population is dominated by young systems. 
Having calculated the orbital evolution of the systems under the
assumption of two different prescriptions for the angular momentum
loss, we find that most of the systems have not yet completed a
significant fraction of their PCEB life time. We therefore predict the
existence of a large population of old PCEBs containing cold white dwarfs
($T_{\mathrm{eff}}\lappr15\,000$\,K).
Our calculations show that nearly half of the PCEBs in our sample
will evolve into a semi-detached configuration and start mass transfer
in less than the Hubble-time. These systems are thus
representative for progenitors of the current CV population. Only one
of them (V471\,Tau) will evolve into a long-period ($\Porb\ga4$\,h)
CV, and a significant fraction of the systems will start mass
transfer in the period gap.
Having estimated the distances of the PCEBs in the sample, we derive a
space density of
$\rho_{\mathrm{PCEB}}\sim6-30\times10^{-6}$\,pc$^{-3}$, depending on
the assumed angular momentum loss prescription.  Taking into account
the evolutionary time scales we compute a lower limit for the CV space
density, predicted by the currently known PCEB population of
$\rho_{\mathrm{CV}}\ga10^{-5}$\,pc$^{-3}$.
Finally, we discuss possible observational selection effects and
conclude that the observed PCEB population is probably highly
incomplete.
\keywords{accretion, accretion discs - binaries: close - novae,
cataclysmic variables.}}
 
\maketitle

\section{Introduction}
In {\em cataclysmic variables} (CVs) \citep[][ for an encyclopaedic
review]{warner95-1} a white dwarf accretes material from its
Roche-lobe filling secondary, typically a K- or M-dwarf.

The standard picture for the \textit{formation} of CVs assumes that
the progenitor systems were moderately wide binaries consisting of an
intermediate mass star with a low mass companion. Once the more
massive star evolves to giant dimensions and fills its Roche-lobe,
runaway mass transfer onto the less massive star starts and the
systems enters into a Common Envelope phase (CE). Friction within the
envelope extracts angular momentum, which tightens the orbit. Post
common envelope binaries (PCEB) with orbital periods of a few days or
less may evolve into a semi-detached CV configuration through orbital
angular momentum loss. Plausible angular momentum loss agencies
are gravitational radiation and --~much more efficient~--
magnetic braking.

The standard paradigm describing the evolution of CVs \textit{after}
the onset of mass transfer is known as \textit{disrupted magnetic
braking} (\citealt{rappaportetal83-1, paczynski+sienkiewicz83-1,
spruit+ritter83-1}; see \citealt{king88-1} for a review).  In brief,
the concept of this theory is that the evolution of CVs is divided
into two main phases, depending on the prevailing angular momentum
loss mechanism. Stellar magnetic braking dominates in CVs whose
Roche-lobe filling donor stars still have a radiative core, which is
the case for orbital periods $\Porb\ge3$\,h. Once that the donor stars
become fully convective at $\Porb\simeq3$\,h, magnetic braking
ceases. For $\Porb<3$\,h gravitational radiation takes over as a much
less efficient angular momentum loss mechanism, resulting in longer
evolution time scales.  As a consequence of the high mass loss rate in
the magnetic braking regime, the donor stars in CVs with $\Porb>3$\,h
are somewhat expanded. The mass loss rate decreases when magnetic
braking ceases at $\Porb\simeq3$\,h, and the donor star reacts by
relaxing to its thermal equilibrium configuration with a radius that
is smaller than its Roche-lobe radius. As a result, the mass transfer
shuts off completely, and the CV becomes an inactive detached
white dwarf/dM
binary that evolves towards shorter periods through emission of
gravitational radiation. At $\Porb\simeq2$\,h, the secondary fills
again its Roche volume and re-starts the mass transfer (though at a
much lower rate than in the long-period CVs).

The main merit of the disrupted magnetic braking model is that it can
successfully explain the period gap, i.e. the statistically
significant paucity of known CVs with orbital periods in the range
$2-3$\,h. However, a number of predictions of this standard model are
in strong disagreement with the observations: (1) the predicted
minimum orbital period is $65-70$\,min, 10\% shorter than the observed
value \citep{kolb+baraffe99-1}; (2) CVs should spend most of their
lifetime near the minimum period, increasing the discovery probability
for such systems. Population syntheses predict a significant
accumulation of systems near the minimum period, which is not observed
\citep{kolb+baraffe99-1}; (3) whereas all population syntheses predict
that $\sim95$\% of the entire CV population should have orbital
periods $<2$\,h \citep[e.g.][]{howelletal97-1}, similar numbers of CVs
are observed above and below the period gap; (4) while the population
models predict a space density of a few $10^{-5}$ to a few
$10^{-4}\,\mathrm{pc^{-3}}$ \citep{dekool92-1, politano96-1}, the
space density derived from the currently known sample of CVs is only
several few $10^{-6}\,\mathrm{pc^{-3}}$
\citep[e.g.][]{downes86-1,ringwald96-1}. If the population models are
correct, we have identified so far only a small fraction
($\sim1-10$\,\%) of the existing CV population
\citep{gaensickeetal02-2}; and finally (5) there is no observational
evidence for a discontinuous change in the spin-down rate due to
magnetic braking between late-type field stars that are fully
convective and those that have a radiative core.

It is apparent that detailed populations studies of the CV progenitors
(i.e. PCEBs) are extremely important for a global understanding of CV
formation and evolution, and extensive theoretical analyses of these
systems have been performed in the past \citep{ritter86-2, dekool92-1,
dekool+ritter93-1, kingetal94-1,politano96-1}. However, the relatively
small number of known PCEBs limited so far the comparison of these
studies with observations. Throughout the last decade, a number of
additional PCEBs have been discovered, significantly improving the
statistical properties of the known PCEB population. 

In this paper, we analyse the properties of a sample of
well-observed PCEBs and discuss possible implications for the PCEB/CV
evolution. In Sect.\,2, we briefly summarise recent
alternatives/additions to the standard CV evolution theory outlined in
the Introduction. The different angular momentum loss prescription
that have been used in the context of CV evolution are described in
Sect.\,3. We introduce our sample of PCEBs in Sect.\,4, and discuss
the past and future evolution of these stars in Sect.\,5 and 6,
respectively. Section\,7 provides the distances to the PCEBs in our
sample.  In Sect.\,8, we compute the space density of PCEBs, and use this
result to estimate the space density of the present-day CV
population. The effects of observational biases are discussed in
Sect.\,9. Finally, Sect.\,10 summarises our findings.

When discussing the properties of the present-day PCEB
population in the context of the present-day CV population,
specifically in Sect.\,6 and 8, we will use the term
\textit{pre-cataclysmic variables} (pre-CVs) to denote those systems
which can be regarded as representative for the progenitors of the
current CV population~--~i.e. PCEBs that evolve into a semi-detached
configuration in less than the Hubble-time\footnote{Obviously also
PCEBs with evolutionary time scales longer than the Hubble-time will
form CVs, but they differ from the progenitors of present-day
CVs. This difference is the only reason for our definition of
``pre-CVs''.}.

\section{Recent modifications of the ``standard scenario'' of CV
formation and evolution} 
The ``standard scenario'' of CV formation and
evolution outlined in the introduction has essentially remained
unchanged during the last two decades, even though being challenged
with a number of alternative suggestions \citep{livio+pringle94-1,
king+kolb95-1, clemensetal98-1, kolbetal98-1}. Very recently, two
far-reaching modifications for the standard scenario have been
proposed.

\citet{sillsetal00-1} and \citet{pinsonneaultetal02-1} presented
theoretical models of the angular momentum evolution of 
low mass stars ($0.1-0.5$\,\Msun) and compared their models to
rotational data from open clusters of different ages to infer
the rotational history of low-mass stars and the dependence of initial
conditions and rotational evolution on mass. 
The studies of \citet{sillsetal00-1}, \citet{pinsonneaultetal02-1}, and 
\citet{andronovetal03-1} 
have two important consequences for the theory of CV evolution. On one
hand, angular momentum loss via magnetic braking is less efficient
above the fully convective boundary than in the standard CV model, and
as a result the evolution time scale of PCEBs containing a
secondary with a radiative core is expected to be 2 orders of
magnitude longer than in the standard model. On the other hand, the
observed angular momentum loss properties show no evidence for a
change in behaviour at the fully convective boundary.  Magnetic
braking remains, hence, an important angular momentum loss mechanism
in PCEBs with a fully convective secondary, and these systems evolve
faster than in the standard scenario.
Due to the increased angular momentum loss below the gap, the
theoretical orbital minimum increases, which is in better agreement
with the observations \citep{kolb+baraffe99-1,patterson98-1,kingetal02-2}.

However, the evolution time scale of CVs above the gap is much longer
than in the standard scenario, and the existence of the observed
period gap is tentatively explained by \cite{pinsonneaultetal02-1} by
two separated populations of CVs with differently evolved donor stars.
One problem with this modification of the standard CV evolution
scenario is that the reduced angular momentum loss above the period
gap predicts mass transfer rates that are significantly lower than the
values derived from the observations. This issue is discussed in more
detail in  Sect.\,\ref{s-aml}.

\citet{king+schenker02-1} and \citet{schenker+king02-1}
\textit{postulate} that the PCEB evolution into a semi-detached
configuration takes much longer than in the standard scenario.
Without modifying the standard angular momentum loss prescription,
such a situation would arise if the frictional angular momentum loss
during the CE is less efficient, the PCEBs exit the CE with wider
orbits, and evolve into the semi-detached CV configuration only on
longer time scales than typically assumed. If the time scale to
initiate mass transfer is of the order of the galactic age
($\sim\,10^{10}$\,yrs), both the minimum orbital period problem
and the space density problem are solved as a much smaller number of
CVs has been produced to date compared to the standard scenario, and
the present-day CV population has not yet reached the minimum period.
Finally \citet{king+schenker02-1} speculate that the longer evolution
time scale for PCEBs proposed in their model might be related to a
reduced magnetic braking as suggested by
\citet{andronovetal03-1}. 
However, if the assumed birth rate of the PCEB progenitor binaries
remains unchanged, a longer PCEB phase \textit{must} result in a very
large population of old PCEBs. In the following sections, we analyse
the properties of the known PCEB sample and discuss the results in the
framework of both the standard CV evolution and the context of a
revised magnetic braking prescription.

\section{Angular momentum loss\label{s-aml}}
Angular momentum loss (AML) drives the evolution of the binaries
and hence, our understanding of AML is \textit{the} main ingredient of
CV and PCEB evolution. It is believed that PCEBs as well as CVs lose
angular momentum via gravitational radiation and magnetic braking.

The rate of AML due to radiation of gravitational waves from
Einstein's quadrupole formula is
\begin{equation}
\dot{J}_{\mathrm{GR}}=-\frac{32\,G^{7/3}}{5c^5}\frac{\Mwd^2\Msec^2}{(\Mwd+\Msec)^{2/3}}\,\left(\frac{2\pi}{\Porb}\right)^{7/3},
\end{equation}
with $\Mwd$ and $\Msec$ the mass of the primary and the secondary.

Unfortunately, the efficiency of magnetic braking is rather uncertain
(see the Introduction and Sect.\,2) and it may have been overestimated
in the past \citep[see][]{pinsonneaultetal02-1}. In the following
analysis of the currently known PCEB sample we use, therefore, two
different prescriptions of AML to calculate the evolution of our
sample of PCEBs.

(1) Disrupted magnetic braking until the secondary becomes fully
convective and magnetic braking ceases \citep[following the standard
scenario of][]{verbunt+zwaan81-1,rappaportetal83-1}. In this case, the
AML is given by
\begin{equation}\label{eq-oldmag}
\dot{J}_{\mathrm{VZ}}=-3.8\times\,10^{-30}\Msec\Rsun^4\left(\frac{\Rsec}{\Rsun}\right)^{\gamma}\left(\frac{2\pi}{\Porb}\right)^{3},
\end{equation}
The original version of the magnetic braking law corresponds to
$\gamma=4.0$ \citep{verbunt+zwaan81-1}. In the context of the CV
evolution $\gamma=2.0$ is frequently used, and hence we adopt this
value throughout this work.

(2) The empirical AML prescription derived from open
cluster data of single stars is \citep{sillsetal00-1}, 
\begin{equation}
\dot{J}_{\mathrm{SPT}}=\left\{
\begin{array}{rr}
-K_{\mathrm{W}}\left(\frac{\Rsec\Msun}{\Msec\Rsun}\right)^{0.5}\omega^3\hspace{1.0cm}\mbox{for}\,\omega\leq\,\omega_{\mathrm{crit}}\,\\
-K_{\mathrm{W}}\left(\frac{\Rsec\Msun}{\Msec\Rsun}\right)^{0.5}\omega\,\omega_{\mathrm{crit}}^2\hspace{0.5cm}\mbox{for}\,\omega>\,\omega_{\mathrm{crit}},
\end{array}\right.
\end{equation}
where $\Ocri$ depends on the mass of the secondary. For stars with
masses $M\,\gappr\,\Msun$ a Rossby scaling matches the observations
\begin{equation}
\Ocri=\Ocri_{\odot}\frac{\tau_{\odot}}{\tau(\Msec)},
\end{equation}
with $\tau$ being the convective overturn time
\citep{krishnamurthietal97-1}.  For lower masses the dependence of
$\Ocri$ on mass is found to be stronger than given by Eq.\,(4)
\citep{sillsetal00-1}. Throughout this work we adopted the values for
$\Ocri$ given by \citet{andronovetal03-1}.  Notice, for the systems in
our sample we find $\omega=10-10^2\times\Ocri $.

\begin{figure}
\includegraphics[angle=-90,width=8.8cm]{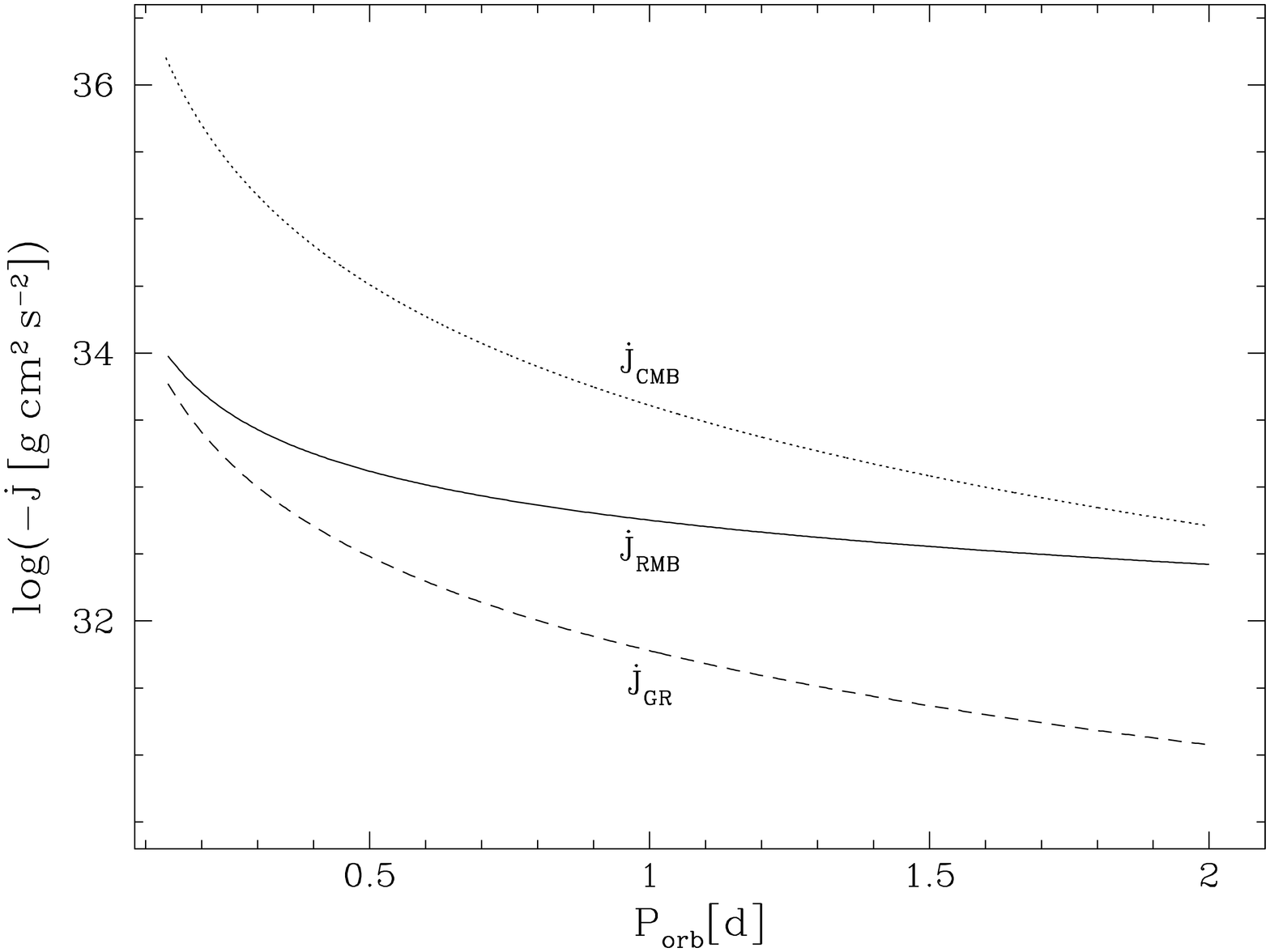}
\caption{\label{f-amlpceb}
Angular momentum loss in PCEBs as a function of the orbital period.
We assumed $\Msec=0.4\Msun$ and $\Mwd=0.6\,\Msun$. Apparently the revised
magnetic braking prescriptions is of similar efficiency as the classical
prescription for long orbital periods whereas it is only a few times the
angular momentum loss due to gravitational radiation for short orbital
periods.} 

\includegraphics[angle=-90,width=8.8cm]{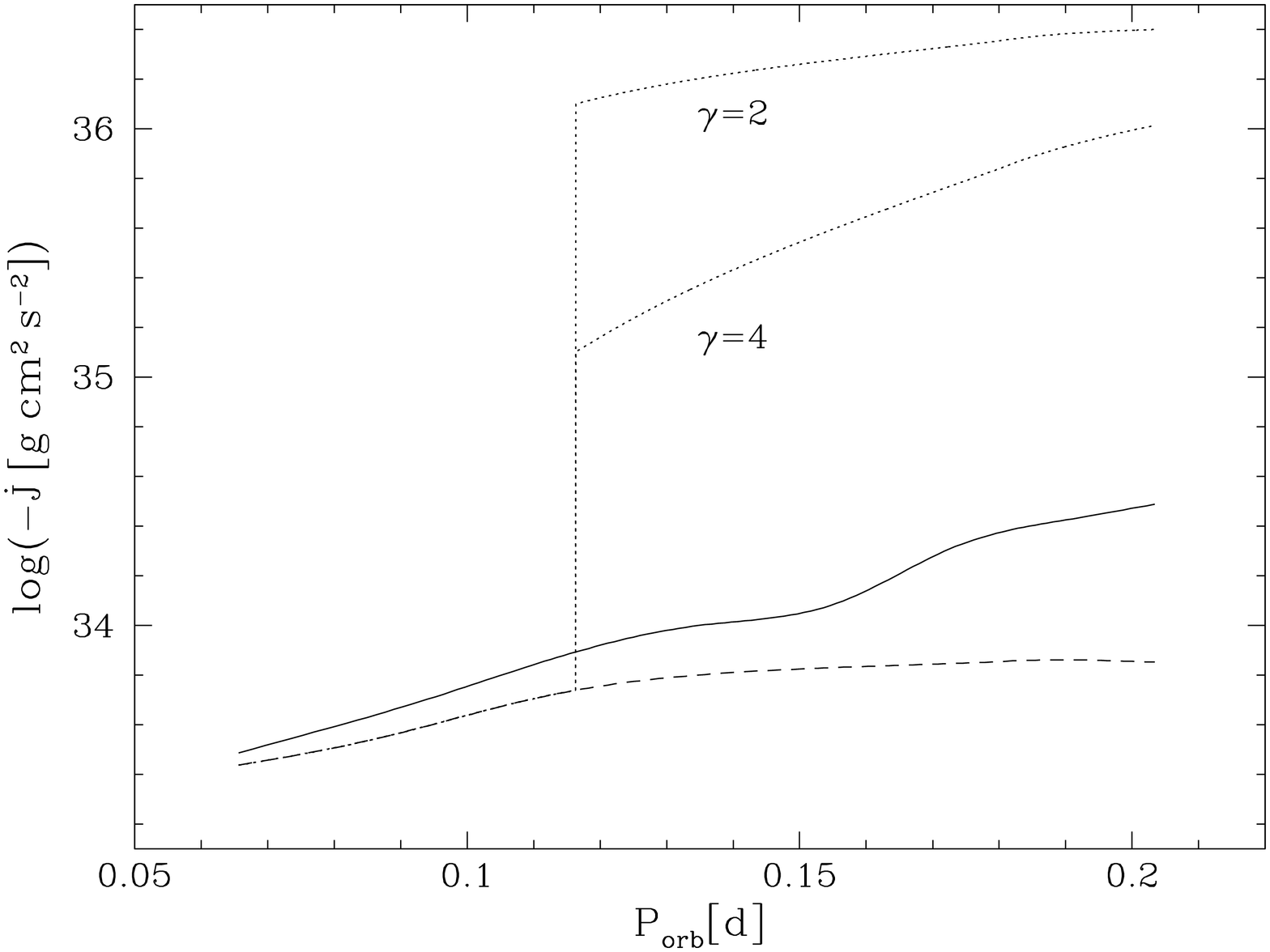}
\caption{\label{f-amlcv}
Angular momentum loss in CVs. As in Fig.\,\ref{f-amlpceb} we used
$\Mwd=0.6\Msun$ \citep[see also][]{pinsonneaultetal02-1,andronovetal03-1}.
The dotted lines correspond to Eq.\,(\ref{eq-jcmb}) with two
values for $\gamma$ (see Eq.\,(\ref{eq-oldmag})). The abrupt decrease
displays the shutoff of magnetic braking for $\Msec<0.3\Msun$. 
The dashed line represents angular momentum loss by gravitational radiation
only whereas the solid line refers to the revised angular momentum loss 
prescription (Eq.\,(\ref{eq-jrmb})). } 
\end{figure}

In the following sections we distinguish between the two currently discussed 
AML prescriptions using the notation
$\dot{J}_{\mathrm{CMB}}$ and $\dot{J}_{\mathrm{RMB}}$ when
referring to the classical respectively the reduced AML prescription:  
\begin{equation}\label{eq-jcmb}
\dot{J}_{\mathrm{CMB}}=\left\{\begin{array}{rr}
\dot{J}_{\mathrm{VZ}}+\dot{J}_{\mathrm{GR}} \hspace{0.5cm}\mbox{for}\hspace{0.5cm} \Msec>0.3\Msun\\
\dot{J}_{\mathrm{GR}} \hspace{0.5cm}\mbox{for}\hspace{0.5cm} \Msec\leq0.3\Msun
\end{array}\right.
\end{equation}
and
\begin{equation}\label{eq-jrmb}
\dot{J}_{\mathrm{RMB}}=\dot{J}_{\mathrm{SPT}}+\dot{J}_{\mathrm{GR}}.
\end{equation}
Fig.\,\ref{f-amlpceb} shows the AML as a function of $\Porb$ for PCEBs. 
Fig.\,\ref{f-amlcv}  compares the angular momentum loss agencies
$\dot{J}_{\mathrm{CMB}}$, $\dot{J}_{\mathrm{GR}}$,
and $\dot{J}_{\mathrm{RMB}}$ for CVs \citep[see
also][]{pinsonneaultetal02-1,andronovetal03-1}.  

It is worth noting that the reduced magnetic braking prescription is
not only subject to observational uncertainties (as stated by
\citet{andronovetal03-1} themselves) but also unable to explain the
mass accretion rates derived from observations for CVs above the
orbital period gap \citep[e.g.][]{patterson84-1}.  This is a firm
conclusion unless (1) all nova-like systems represent short-lived
high accretion states or (2) there exists an additional AML mechanism in these
systems. Referring to (1) we note, the possibility of irradiation induced mass 
transfer cycles \citep[e.g.][]{kingetal95-1} 
and that nova eruptions widen the mass transfer spectrum \citep{kolb02-1}. 
Concerning (2) we note that circumbinary
disks in CVs have been suggested as rather efficient additional AML
agents \citep[][]{spruit+taam01-1,dubusetal02-1}.
Thus, while the reduced AML prescription has a
problem to explain the ``observed'' mass transfer rates of CVs  
above the gap, it is currently not possible to rule out its 
validity because of the additional complications just mentioned
\footnote{In this context we note that it might be easier to test our 
understanding of AML using a large (not yet identified) sample of PCEBs, 
as all the complications related to mass transfer are not present in these
binaries.}.

Considering the described uncertainties of the AML in close binaries, we 
discuss the current PCEB population in the context of both
prescriptions, \Jcmb\ and \Jrmb.

\section{PCEBs: The present}

\begin{figure}
\includegraphics[angle=-90,width=8.8cm]{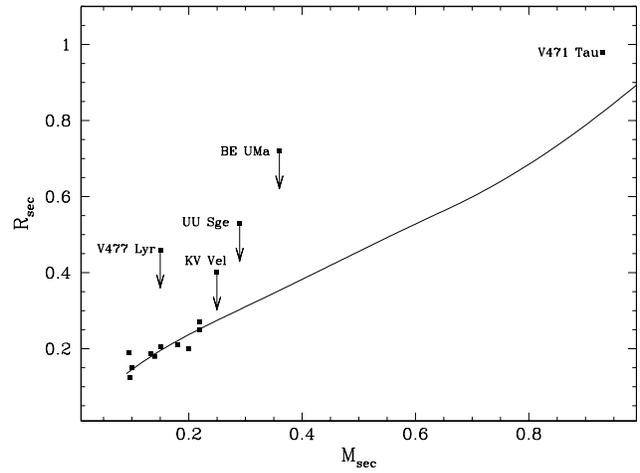}

\caption{\label{f-secrad}
The ZAMS mass radius relation given by \citet{politano96-1} and the
positions of the PCEB secondaries for which an observationally determined
radius is available. The arrows indicate that the radii of the secondaries in
these planetary nebulae are larger due to recent accretion onto the secondary
during a CE phase. Before becoming a CV the secondary will have contracted to
its main-sequence radius. }
\end{figure}

\begin{table*}
\newcounter{ref}
\newcommand{\tcite}{\stepcounter{ref}\arabic{ref}}
\newcommand{\tref}[1]{\stepcounter{ref}(\arabic{ref})\,\citealt{#1}}
\caption[]{\label{t-sample}Sample of PCEBs with orbital periods
$P\lappr\,2$\,days, $\Mwd>\Msec$ and a main-sequence secondary star.}
\setlength{\tabcolsep}{1.01ex}
\begin{tabular}{lllllllll}
\hline\noalign{\smallskip}
Object&
alt. name&
Sp1 &
\multicolumn{1}{c}{\Porb\,[d]}&
\multicolumn{1}{c}{\Mwd\,$[M_{\odot}]$}&
\multicolumn{1}{c}{\Msec\,$[M_{\odot}]$} &
\multicolumn{1}{c}{\Rsec\,$[R_{\odot}]$} &
\multicolumn{1}{c}{\Twd\,[K]}&
Ref.\\
\noalign{\smallskip}\hline\noalign{\smallskip}
RR\,Cae  & LFT 349             & WD &
 0.304&$0.467$      &0.095 & 0.189 & $7000 $& \tcite,\tcite\\
EG\,UMa  & Case 1              & WD & 
 0.668 & $0.64\pm0.03$ & $0.42\pm0.04$ & $0.40^{\%}$ &
$13125\pm125$&\tcite\\
EC\,13471--1258&                 & WD & 
 0.151 & $0.77\pm0.04$ & $0.58\pm0.05$ & $0.42$ &    
$14085\pm100$&\tcite\\ 
BPM\,71214 &                   & WD & 
 0.201 &  $0.77\pm0.06$ & 0.4           & $0.38^{\%}$ & 
$17200\pm1000$&$4$\\ 
HR\,Cam  & GD\,448             & WD &
 0.103&$0.41\pm0.01$&$0.096\pm0.004$&$0.125\pm0.020$ & $19000$& \tcite,\tcite\\
UZ\,Sex  & PG\,1026+002        & WD &
 0.597&$0.65\pm0.23$&$0.22\pm0.05$&$0.25\pm0.03$ & $17600\pm2000$&\tcite\\
& & & & & & &$19900\pm330$&\tcite\\ 
BPM\,6502& LTT\,3943           & WD &
 0.337&$0.5\pm0.05$&$0.16\pm0.09$& $0.204^{\%}$ &$20640\pm200$&\tcite\\
& & & & & & &$21380\pm258$& $2$\\ 
& & & & & & &$20311\pm532$&\tcite\\  
HZ\,9    &                     & WD & 
 0.564&$0.51\pm0.1$&$0.28\pm0.04$& $0.296^{\%}$ &$20000\pm2000$& \tcite,\tcite,\tcite\\
& & & & & & &$17400$& this work\\  
MS\,Peg  & GD\,245             & WD &
 0.174&$0.48\pm0.02$&$0.22\pm0.02$& $0.27\pm0.02$&$22170 $&\tcite\\
CC\,Cet  & PG 0308+096         & WD &
 0.284&$0.39\pm0.1$ &$0.18\pm0.05$ & $0.21\pm0.02$ & $26200\pm2000$ & $7$\\
HW\,Vir&BD$-07^{\circ}4377$    & sdB &
 0.117&$0.48\pm0.09$&$0.14\pm0.02$&$0.180\pm0.011$&$28488\pm208 $&\tcite\\
HS\,0705+6700 &                & sdB &
 0.096 & 0.483 & 0.134 & $0.186$ & 28800 & \tcite\\
LM\,Com  & Ton\,617$^\dag$        & WD &
 0.259&$0.45\pm0.05$&$0.28\pm0.05$& $0.296^{\%}$ & $29300$& \tcite\\
PG\,1017--086 &                & sdB &
 0.073 & 0.5 & $0.078^{+0.005}_{-0.006}$& $0.085\pm0.04$ &30300& \tcite\\
V471\,Tau&BD$+16^{\circ}516$   & WD &
 0.521&$0.84\pm0.05$&$0.93\pm0.07$&$0.98\pm0.10$&$34500\pm1000$& \tcite\\
NY Vir       &  PG\,1336--018 & sdB &
 0.101 & 0.5 & 0.15 &$0.205\pm0.01$& $33000\pm3000$ &\tcite\\
AA\,Dor  & LB\,3459            & sdO &
 0.262&$0.330\pm0.003$&$0.066\pm0.001$& $0.10\pm0.01^{!}$ & $42000\pm1000$ &\tcite\\
RE\,2013+400&                  & WD &
 0.706&$0.56\pm0.03$&$0.18\pm0.04$& $0.221^{\%}$ &$49000\pm700$ &\tcite\\
& & & & & & &$47800\pm2400$&\tcite\\  
GK\,Vir & PG\,1413+015         & WD &
 0.344&$0.51\pm0.04$&0.1 & $0.15$ & $48800\pm1200$ &\tcite,\tcite\\
MT\,Ser  & PN A66 41$^*$       & sdO,PN &
 0.113&$0.6 $&$0.2\pm0.1$ & $0.2\pm0.1$ & $50000 $& \tcite,\tcite\\
IN\,CMa&RE\,J0720--318         & WD &
 1.26&$0.57\pm0.02$&$0.39\pm0.07$&$0.375^{\%}$ &$53000\pm1100$& $22$\\
NN\,Ser&PG\,1550+131           & WD &
 0.130&$0.57\pm0.04$&$0.12\pm0.03$&$0.166^{\%}$&$55000\pm8000$& \tcite\\ 
TW\,Crv  &  EC\,11575--1845 & sdO &
 0.328 & & & & $95000\pm40000$ & \tcite\\
RE\,1016--053&                 & WD &
 0.789&$0.6\pm0.02$&$0.15\pm0.02$& $0.195^{\%}$ &$55000\pm1000$&$22$\\
& & & & & & &$56400\pm1200$&$23$\\  
UU\,Sge&PN A66 63$^*$          & sdO,PN &
 0.465&$0.63\pm0.06$&$0.29\pm0.03$& $0.53\pm0.02$ & $87000\pm13000$& \tcite,\tcite,\tcite\\
& & & & & & &$57000\pm8000$&$32$\\ 
V477\,Lyr & PN A66 46$^*$& sdOB,PN &
  0.472 & $0.51\pm0.07$ & $0.15\pm0.02$ & $0.46\pm0.03$ & 60000 & \tcite\\
PN A66 65    & Abell 65        & sd?,PN &
 1.00& & & &$80000 $&\tcite,\tcite\\
KV\,Vel  & LS\,2018            & sdO,PN &
 0.357&$0.63\pm0.03$&$0.25\pm0.06$& $0.402\pm0.005$ & $77000\pm3000$ & \tcite\\
& & & & & & & 90000 & \tcite\\
HS 1136+6646   &             & WD &
 1 & & & & 100000 &\tcite\\
BE\,UMa&PG\,1155+492           & sdO,PN &
 2.291&$0.7\pm0.07$&$0.36\pm0.07$&$0.72\pm0.05$ & $105000\pm5000$& \tcite,\tcite\\
\noalign{\smallskip}\hline\noalign{\smallskip}
\end{tabular}
\linebreak
Other frequently used designations: $^*$
Abell\,41, 46 \& 63 (to be used with care as the object numbers in
\citealt{abell55-1} and \citealt{abell66-1} are not consistent); $^{\dag}$
PG\,1224+309. \\
$^{!}$ The radius of the secondary is
estimated using the mass-radius relation shown by \citet{kudritzkietal82-1}
(their Fig.\,8).\\
$^{\%}$ Radii of the secondary calculated
by using the mass-radius relation given by \citet{politano96-1}.\\
\setcounter{ref}{0}
References: 
\tref{bruch99-1}, \tref{bragagliaetal95-1},
\tref{bleachetal00-1},
\tref{kawkaetal02-1},
\tref{marsh+duck96-1}, \tref{maxtedetal98-1},
\tref{safferetal93-1}, \tref{kepler+nelan93-1}, 
\tref{kawkaetal00-1}, \tref{koesteretal79-1},  
\tref{stauffer87-1}, \tref{lanning+pesch81-1},\tref{guinan+sion84-1}
\tref{schmidtetal95-3}, 
%
\tref{wood+saffer99-1},
\tref{drechseletal01-1},
\tref{oroszetal99-1},
\tref{maxtedetal02-1},
\tref{obrienetal01-1},
\tref{kilkennyetal98-1},
\tref{rauch00-1},
\tref{vennesetal99-2}, \tref{bergeronetal94-1},
\tref{fulbrightetal93-1}, \tref{greenetal78-1},
\tref{greenetal84-1}, \tref{grauer+bond83-1},
%
\tref{catalanetal94-1},
\tref{chenetal95-1}
%
\tref{milleretal76-1}, \tref{pollacco+bell93-1}, \tref{belletal94-1},
\tref{pollacco+bell94-1},
\tref{bond+livio90-1}, \tref{walsh+walton96-1},
\tref{hilditchetal96-1}, \tref{herreroetal90-1}, 
\tref{singetal01-1},
\tref{woodetal95-3}, \tref{fergusonetal99-1}.
\end{table*}

We have compiled for our present study a sample of PCEBs with
well-established characteristics from various literature sources,
including the catalogues of \citet{ritter+kolb98-1} and
\citet{kubeetal02-1}. We exclude from our sample systems with
$\Msec>\Mwd$ (such as V651\,Mon), systems with $\Porb\ga2$\,d (such as
Feige 24), and systems with sub-giant secondary stars (such as
FF\,Aqr). Table\,1 lists the properties of our PCEB sample, which
consists of 18 systems containing a white dwarf primary and 12 systems
containing an sdOB primary (of which four are the central stars of
planetary nebulae).  These 30 systems represent the currently known
population of well-observed PCEBs which will evolve eventually into a
semi-detached configuration and turn into a CV and for which reliable
observational data exists.

It is interesting to note that the radii of the secondaries derived
from observations (Table\,1) are generally in good agreement with the
ZAMS mass-radius relation given by \citet{politano96-1}. Exceptions
are the secondary stars in the binary planetary nebulae BE\,UMa,
KV\,Vel, V477\,Lyr, and UU\,Sge whose radii exceed the ZAMS value by a
factor of $\sim\,2$ (Fig.\,\ref{f-secrad}).  This effect can be
explained by accretion of matter onto the secondary at a
relatively high rate during the recent CE-phase. Numerical
calculations of accreting low-mass stars have been performed by
\citet{prialnik+livio85-1} for a fully convective star of $M=0.3\Msun$
and \citet{fujimoto+iben89-1} for a low mass main sequence star of
$M=0.75\Msun$. The calculations show that rather recent accretion at a
rate of $\sim\,10^{-5}\,\Msun\,\mathrm{yrs}^{-1}$ to
$\sim\,10^{-7}\,\Msun\,\mathrm{yrs}^{-1}$ can explain the observed
expansion of the secondary.  As the time scale on which the secondary
contracts is shorter than the binary evolution time scale we use radii
calculated from the mass-radius relation of \citet{politano96-1} for
the four planetary nebulae in our calculations (Sect.\,\ref{s-past},
and \ref{s-future}).

A peculiar system is the Hyades PCEB V471\,Tau, located in the upper
right corner of Fig.\,\ref{f-secrad}. Apparently, the secondary is
expanded with respect to the ZAMS mass radius relation. Using
radiometric measurements and ground based Doppler imaging
\citet{obrienetal01-1} explain the oversized secondary in V471\,Tau as
the result of a large fraction of the secondaries surface being
covered by star spots, lowering its average effective temperature. In
order to maintain the luminosity of a 0.93\,$\Msun$ dwarf, the radius
of the star has, hence, to exceed the ZAMS value.  It is worth to note
that also the white dwarf in V471\,Tau represents an evolutionary
paradox as it is the hottest (and therefore youngest) as well as the
most massive white dwarf in the Hyades cluster. \citet{obrienetal01-1}
discuss the possibility that V471\,Tau is descended from a blue
straggler, which would imply that its progenitor was initially a
triple system.  Nevertheless, despite its curious nature, V471\,Tau is
the best-observed PCEB, and, as described in detail in
Sect.\,\ref{s-future}, it will be the PCEB from our sample that starts
mass transfer at the longest orbital period.

Final notes concern MT\,Ser, UU\,Sge, 
EC\,13471--1258, BPM\,71214, PG\,1017--086
and AA\,Dor.  The orbital period
of MT\,Ser might be as twice as long as given in Table\,1, and the
system may consist of two hot subdwarfs rather than a sub\-dwarf plus
cool companion \citep{bruchetal01-1}. The large discrepancy between the
two values for the temperatures of the primary in UU\,Sge given in
Table\,1 results from the fact that \citet{belletal94-1} could fit the
observed light curve equally well by assuming either limb-darkening or
limb-brightening. The given values may therefore be interpreted as
upper and lower limits. 
The binary parameter for EC\,13471--1258 and BPM\,71214 given in Table\,1
indicate these systems are extremely close to the semi-detached state. 
Therefore \citet{kawkaetal02-1} suggested that at least EC\,13471--1258 
might also be a hibernating nova instead of a PCEB. 
We additionally note that the radius obtained for the secondary in 
EC\,13471--1258 is essentially lower than predicted by the main-sequence 
mass radius relation which appears implausible and indicates
rather large uncertainties in the values for $\Msec$ and/or $\Rsec$.  
Finally, the classification of the secondaries in
PG\,1017--086 and AA\,Dor is not unambiguous. It has been noted 
that in both systems the secondary may also be a brown dwarf rather than a 
low mass M-dwarf \citep{maxtedetal02-1,rauch00-1}.

\begin{figure*}
\begin{center}
\includegraphics[width=7.4cm, angle=270]{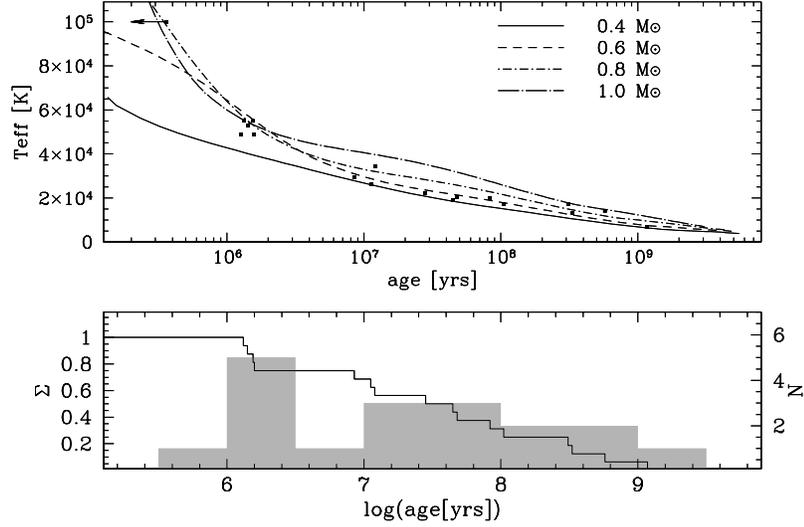}
\caption[]{\label{f-coolingtracks}
Cooling tracks for non-accreting white dwarfs after
\citet{wood95-1} for different
masses of the white dwarf.  The position of the currently known
PCEBs (Table 1) are obtained by interpolating between the cooling
tracks for different masses of the primary.  The bottom panel shows
the age of the PCEB population in a histogram with logarithmic bins
(shaded, N) and the relative cumulative distribution (solid line,
$\Sigma$).}
\end{center}
\end{figure*}

\section{PCEBs: The past\label{s-past}}

\subsection{\label{s-coolingage} The white dwarf cooling age}

The cooling of single white dwarfs has been modelled to a very high
degree of precision allowing to derive accurate age estimates for
field white dwarfs from their (observationally determined) effective
temperatures and masses. Some uncertainties in the theoretical models
remain at the low-temperature end of the white dwarf luminosity
function, and are relevant primarily for the cosmo-chronological
interpretation of the coldest (halo) white dwarfs (see the reviews by
\citealt{fontaineetal01-1, koester02-1}).

We determined the cooling ages for the PCEBs in our sample which
contain a white dwarf primary by interpolating Wood's
(\citeyear{wood95-1}) CO core evolution tracks for the effective
temperatures and masses listed in Table\,\ref{t-sample}\footnote{Note
that the same method \textit{can not be applied in CVs}, as accretion
(re)heats the white dwarf. As a result, white dwarfs in CVs are
typically hotter than field white dwarfs of comparable age and mass
\citep{gaensicke00-1,townsley+bildsten02-1}.}.  The cooling age
estimates are given in Table\,\ref{t-future}, and are shown in
Fig.\,\ref{f-coolingtracks} along with Wood's cooling tracks. The
bottom panel of Fig.\,\ref{f-coolingtracks} displays the age
distribution of our PCEB sample. It is evident that the
\textit{currently known} population of PCEBs is strongly dominated by
systems younger than $5\times10^8$\,yrs, with the exception of
RR\,Cae, which has a cooling age of
$t_{\mathrm{cool}}\sim\,10^9$\,yrs.  The mean cooling age of the 18
systems in the sample containing a white dwarf primary is
$\overline{t}_{\mathrm{cool}}\sim\,1.4\times\,10^8$\,yrs.  Considering
that the sdOB, sdO, and PN systems in our sample are white dwarf
progenitors, the mean age of the entire PCEB sample is even lower.
The accuracy of our cooling age estimates depends (a) on the
uncertainties in the white dwarf masses and (b) on the core
composition of the white dwarfs. We discuss both issues in more detail
below.

(a) The white dwarf masses in PCEBs can be measured to a rather high
level of precision, as several independent methods are available in
these systems (fitting model spectra to the Balmer lines, dynamical
measurements from the radial velocity variations of the white dwarf
and its companion, measurement of the gravitational redshift of
photospheric lines from the white dwarf). The errors on the white
dwarf masses extracted from the literature (Table\,\ref{t-sample}) are
completely negligible in the context of our cooling age
estimates. Even if we assume that the published errors are
underestimated, and that the true errors in the white dwarf masses are
of the order $\pm0.1-0.2$\Msun, the corresponding
uncertainties in the derived cooling ages are of the order $\sim2$,
which is still irrelevant for the comparison to the binary evolution
time scales.

(b) The evolution of a white dwarf progenitor in a binary system may
affect the core composition of the white dwarf.  If the first Roche
lobe overflow phase occurs prior to helium ignition in the primary
star \citep{dekool+ritter93-1, iben+tutukov93-1} the resulting white
dwarf will be of low mass and have a helium core.  Consequently,
evolution models for CVs predict that systems with a low mass primary
($\Mwd<0.5\Msun$) should contain a helium core white dwarf
\citep{dekool+ritter93-1, dekool92-1,politano96-1,howelletal01-1}.
Careful observations of seven previously considered low mass single
white dwarfs revealed that five of them are indeed close binary
systems \citep{marshetal95-1}.

Comparing the cooling models for CO white dwarfs \citep{wood95-1} and
He white dwarfs \citep{althaus+benvenuto97-1, driebeetal98-1,
driebeetal99-1} shows, however, that the evolution of CO and He white
dwarfs differs noticeably only for very low-mass stars, where residual
hydrogen shell burning dominates over the gravothermal energy release.
Inspecting Figs.\,2 and 3 of \citet{driebeetal98-1} shows that only for a
single system from our sample (Table\,\ref{t-sample}) the cooling age
derived from Wood's (\citeyear{wood95-1}) CO tracks may be significantly
wrong: if the white dwarf in RR\,Cae $(0.467\,\Msun,
7000\,\mathrm{K})$ contains a He core, then its cooling age is
$\sim4-5\times10^9$yrs, compared to $1.2\times10^9$\,yrs for a CO
core. We conclude that the effect of different core compositions is
negligible in the context of the present paper, but has to be taken
into account in future analyses if additional PCEBs
containing low mass ($\Mwd\la0.4\,\Msun$) and cool
($\Teff\la20000$\,K) white dwarfs are found.

A final note concerns the possible effect of re-accretion of
CE material, which may alter the composition of the white dwarf
envelope in a PCEB with respect to that of a single white dwarf.
Differences in envelope/atmosphere composition will affect the cooling
of the white dwarf only at the low-temperature end of the luminosity
function. While this effect is important for the  cosmochronological
interpretation of the coldest and oldest halo WDs
\citep[e.g.][]{koester02-1}, it is irrelevant for our PCEB
sample.

\subsection{The orbital period at the end of the common envelope phase}

During the detached post CE-phase the masses of the
companions remain essentially constant,
i.e. $\dot{M}_{\mathrm{wd}}=\dot{M}_{\mathrm{sec}}=0$. With Kepler's
third law this leads to the well known relation
\begin{equation}\label{eq-int}
\frac{\dot{J}}{J}=\frac{\dot{P}}{3P}.
\end{equation}
Based on the the cooling age of a given PCEB derived from cooling tracks
(Fig.\,\ref{f-coolingtracks}, Table\,\ref{t-future}) we can
approximate the orbital period at the end of the CE phase by
integrating Eq.\,(\ref{eq-int}).
We have done this calculation for both
the standard scenario (Eq.\,(\ref{eq-jcmb})), and for the empirical AML
prescription by \citet{sillsetal00-1} (Eq.\,(\ref{eq-jrmb})).  In the
standard scenario magnetic braking is assumed to be much stronger than
gravitational radiation for $\Msec>0.3\Msun$ and, hence, we can simply
integrate Eq.\,(\ref{eq-int}) using either
$\dot{J}=\dot{J}_{\mathrm{GR}}$ or $\dot{J}=\dot{J}_{\mathrm{VZ}}$
depending on the mass of the secondary.  For gravitational radiation
only we get:
\begin{eqnarray}\label{eq-sol1}
t_{\mathrm{cool}}&=&\frac{5c^5}{256G^{5/3}(2\pi)^{8/3}}\frac{(\Mwd+\Msec)^{1/3}}{\Mwd\Msec}\nonumber\\
& &\hspace{3cm}\times(P_{CE}^{8/3}-\Porb^{8/3}),
\end{eqnarray}
\citep[see also][]{ritter86-2}.  For systems with $\Msec>0.3\Msun$ we
assume $\dot{J}=\dot{J}_{\mathrm{VZ}}$ and derive:
\begin{eqnarray}\label{eq-sol2}
t_{\mathrm{cool}}&=&\frac{2.63\,10^{29}G^{2/3}\Mwd}{(2\pi)^{10/3}(\Mwd+\Msec)^{1/3}}\Rsun^{-4}\left(\frac{\Rsun}{\Rsec}\right)^{\gamma}\nonumber\\
& &\hspace{3cm}\times(P_{CE}^{10/3}-\Porb^{10/3}).
\end{eqnarray}

Considering the empirical angular momentum loss prescription of
\citet{sillsetal00-1} the situation is somewhat more complex as the
two angular momentum loss agencies can be of the same order of magnitude (see
Fig.\,\ref{f-amlpceb}) and we can not neglect gravitational radiation.  
Nevertheless, there exists also an analytical solution for the integral of
Eq.\,(\ref{eq-int}) using $\dot{J}=\dot{J}_{\mathrm{RMB}}$:
\begin{eqnarray}\label{eq-sol3}
t_{\mathrm{cool}}&=&b{\Biggl[}\frac{3(1+aP^{4/3})}{4a(P^{-5/3}+aP^{-1/3})P^{1/3}}-\nonumber
\\
& & \hspace{0.75cm}{\frac{3(1+aP^{4/3})\ln(1+aP^{4/3})}{4a^2(P^{-5/3}+aP^{-1/3})P^{5/3}}}\Biggr]^{\Porb}_{P_{\mathrm{CE}}},
\end{eqnarray}
with 
\begin{eqnarray}
&b&=\left(\frac{96G^{5/3}\Mwd\Msec(2\pi)^{8/3}}{5c^5(\Mwd+\Msec)^{1/3}}\right)^{-1}\nonumber\\
&a&=\left(\frac{\Msun\Rsec}{\Msec\Rsun}\right)^{0.5}\frac{K_w\Ocri^2(\Mwd+\Msec)^{2/3}5c^5}{32(2\pi)^{4/3}G^{7/3}\Mwd^2\Msec^2}\nonumber
\end{eqnarray}
Having derived the cooling age of the PCEBs we solve
Eqn.\,(\ref{eq-sol1}), (\ref{eq-sol2}), and (\ref{eq-sol3}) to obtain the
orbital period immediately 
after the CE phase 
($P_{\mathrm{CE}}$). Again, the results are given in Table\,\ref{t-future}.  
Not surprisingly, $P_{\mathrm{CE}}$ is very close to the present orbital
period of the systems as we found most of the PCEBs to be quite young.
For systems with $t_{\mathrm{cool}}\lappr\,10^7$\,yrs we get essentially
$\Porb\,\simeq\,P_{\mathrm{CE}}$.  Clearly, the assumed AML prescription makes
a significant difference only for rather old PCEBs (i.e. RR\,Cae, in our
sample) or for systems not too young and with $\Msec>0.3\Msun$ (EG\,UMa,
V471\,Tau, EC\,13471--1258, BPM\,71214).  In these cases we give in
Table\,\ref{t-future} both results  
where the second line corresponds to
$\dot{J}=\dot{J}_{\mathrm{RMB}}$.  Comparing the
obtained distribution of orbital periods after the CE-phase
($P_{\mathrm{CE}}$) with theoretical predictions \citep[Fig.\,9
in][]{dekool92-1}, we find reasonable agreement: for our systems we
get $0.4>\log(P_{\mathrm{CE}})>-1.1$ with $\sim\,60\%$ in the range of
$-0.1>\log(P_{\mathrm{CE}})>-0.6$.  

\subsection{The total age of the PCEBs}

The cooling age of the primary represents just the age of the systems
after the end of the CE. To approximate the total age of the binary,
i.e the time since it appeared on the main sequence we use simple
analytical fits to the stellar evolution, Roche geometry, and Kepler's
third law to obtain the nature of the initial main sequence binary.
Combining Eqn.\,(35,36), and (43)--(49) of \citet[][]{politano96-1}
with the standard prescription of the binary shrinkage during the
CE-phase \citep[Eq.\,(8) in][]{dekool92-1} gives the mass of the
initial primary and, hence, the nuclear evolution time. This is,
of course, just a rough approximation because an uncertain fraction
$\alpha_{\mathrm{CE}}$ of the gravitational binding energy that is
released when the secondary and the primary spiral together goes into
the ejection of the envelope. However, we assume $\alpha_{\mathrm{CE}}=1$
and $\lambda=0.5$ \citep[see][]{dekool92-1} which gives results in
reasonable agreement with the initial-to-final mass relation for
CO-white dwarfs \citep{weidemann00-1}.  The estimates obtained for
the masses of the progenitors of the primaries ($M_{1,\mathrm{pCE}}$)
and the resulting evolution time since the binary appeared on the main
sequence ($t_{\mathrm{evol}}$) are also given in
Table\,\ref{t-future}.  For PCEBs containing a low mass white dwarf
$t_{\mathrm{evol}}$ can be of the order of $10\%$ of the Hubble time
(see Table\,\ref{t-future}) and therefore is not generally negligible.

\section{PCEBs: The future\label{s-future}}

\subsection{How long before the onset of mass transfer?}

In the last section we showed that the vast majority of the known
PCEBs are young objects.  Here, we calculate the orbital periods that
these binaries will have when they turn into CVs as well as the time
scale for this evolution. 

\begin{figure}
\includegraphics[angle=-90,width=8.8cm]{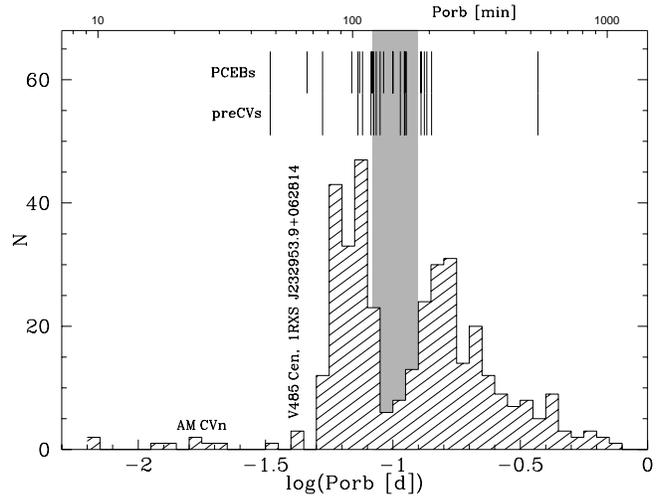}
\caption{\label{f-cvporb} The orbital period distribution of the
currently known CV population (from
\citealt{kubeetal02-1}). Grey-shaded is the orbital period gap. The
vertical lines on the top indicate the periods at which the PCEBs from
Table\,1 will enter a semi-detached configuration and start mass
transfer. Those PCEBs that will evolve into CVs within less than a
Hubble time (Table\,2) are considered to be genuine pre-CVs. 
The two PCEBs with $\Psd$ below the minimum orbital period of CVs at
$\sim\,80$\,min are the two systems in our sample which possibly have a 
brown dwarf secondary AA\,Dor and PG\,1017--086 (see Sect.\,4).}
\end{figure}

\begin{figure}
\includegraphics[angle=0,width=8.5cm]{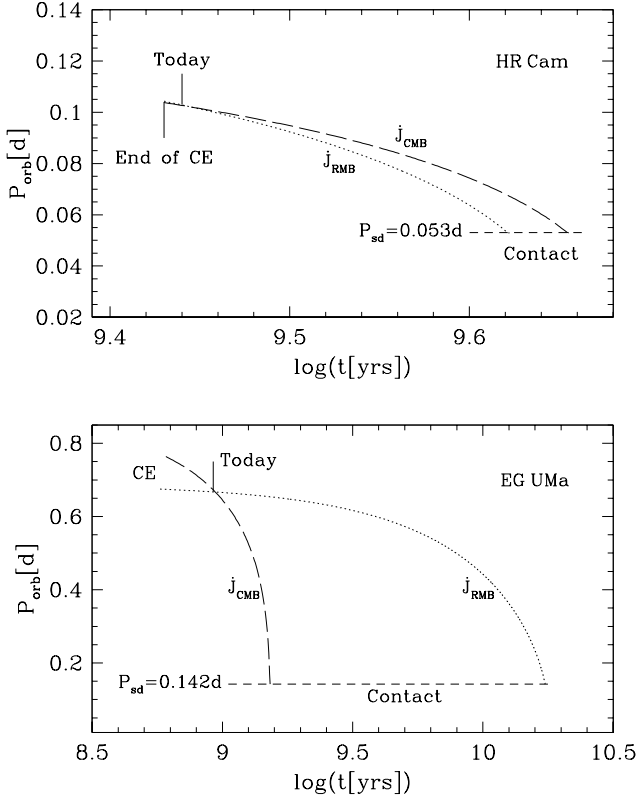}
\caption{\label{f-amlcmp} The PCEB evolution of HR\,Cam and EG\,UMa
assuming two different prescriptions for the AML, \Jcmb\ and \Jrmb. As
the mass of the secondary in HR\,Cam is significantly below $0.3\Msun$
the system evolves slower into contact if we assume AML according to
the standard scenario. Conversely, EG\,UMa turns into a CV much faster
in the standard scenario because of the more efficient magnetic
braking for $\Msec>0.3\Msun$. }
\end{figure}

\begin{figure}
\includegraphics[angle=-90,width=8.5cm]{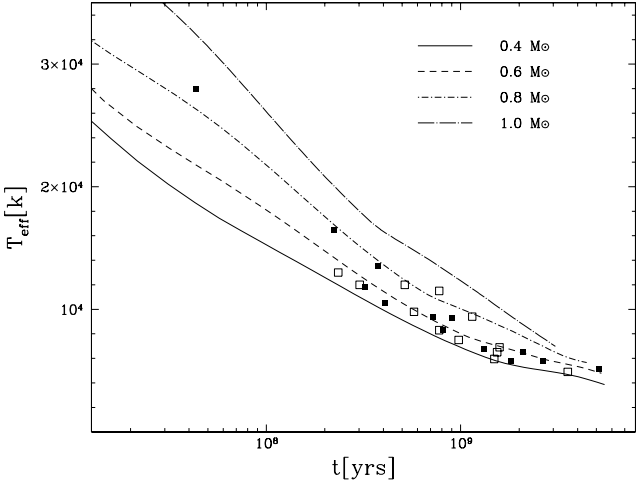}
\caption{\label{f-teffcon}
The cooling tracks of \citet{wood95-1} and the locations in 
the age/temperature plane where mass transfer will
initiate for the PCEBs with a short PCEB lifetime
($t_{\mathrm{cool}}+\tsd<\,6\times\,10^{9}$\,yrs).
The filled squares correspond to $\Jcmb$ the open squares to
$\Jrmb$. Notice, most of the PCEBs will stay more than $10^{10}$\,yrs
in the PCEB state and, hence, will have $\Twd<6000$\,K when becoming
semi-detached.}
\end{figure}

Knowing $\Mwd$ and $\Msec$ of a (detached) PCEB (Table\,1) the
orbital period of the corresponding semi-detached configuration, \Psd\
follows from Roche geometry and Kepler's third law \citep{ritter86-2}:
\begin{equation}
\Psd=9\pi\left(\frac{\Rsec^3}{G\Msec(1+\Mwd/\Msec)(R_\mathrm{L}/a)^3}\right)^{0.5}.
\end{equation}
Here $R_\mathrm{L}$ denotes the volume radius of the Roche-lobe
of the secondary and $a$ the binary separation. We use the
formula of \citet{eggleton83-1} to relate these quantities to the mass
ratio of the system.  The resulting \Psd\ for the PCEBs in our sample
are listed in Table\,\ref{t-future} and are compared to the orbital
period distribution of the currently known CV population in
Fig.\,\ref{f-cvporb}. It is apparent that only a small number of
systems ($\sim\,22\%$) of the currently known PCEB population will evolve into
CVs above the period gap ($\Psd>3$\,h), whereas $\sim\,30\%$ will evolve
into contact at $\Psd<2$\,h. A somewhat surprising result is that
almost  half ($\sim\,48\%$) of the currently known PCEBs will start their
CV life in the period gap ($2$\,h$<\Psd<3$\,h). The small number
of progenitors of long-period CVs will be discussed in
Sect.\,\ref{s-selection}. 

Knowing $\Psd$ we now calculate the time it will take a PCEB to become
a CV ($\tsd$), again by integrating Eq.\,(\ref{eq-int}) assuming either AML
according to the ``standard scenario''
($\dot{J}=\dot{J}_{\mathrm{CMB}}$) or Sill's et
al. (\citeyear{sillsetal00-1}) empiric AML prescription
($\dot{J}=\dot{J}_{\mathrm{RMB}}$). 
The solution for $\tsd$ can be derived by replacing
$t_{\mathrm{cool}},\,\Porb,\,P_{\mathrm{CE}}$ with
$\tsd,\,\Psd,\,\Porb$ in Eqn.\,(\ref{eq-sol1}--\ref{eq-sol2}).  It is
worth to note that although the PCEBs evolve on a time scale
comparable or even longer than $10^{10}$\,yrs, the expansion of the
secondary due to its nuclear evolution is negligible for all the PCEBs
in our sample and, hence, not taken into account in our calculations.
Inspecting the resulting \tsd\,(Table\,\ref{t-future}), it becomes
evident that most of the PCEBs have completed only a small fraction
$f_{\mathrm{PCEB}}$ of their predicted PCEB-lifetime.  On average the
total evolution time for the PCEBs is $\sim3.4\times10^{10}$\,yrs
independent on the assumed AML prescription\footnote{Notice, \Jrmb\ is
smaller than \Jcmb\ only above the gap otherwise \Jrmb is more
efficient.}.

Assuming $\dot{J}=\dot{J}_{\mathrm{CMB}}$, four systems (i.e. EG\,UMa,
EC\,13471--1258, BPM\,71214, V471\,Tau) have $f_{\mathrm{PCEB}}>0.1$.
EC\,13471--1258 is expected to be very close to the onset of mass
transfer, $f_{\mathrm{PCEB}}>0.995$, and one may speculate that this
system is an old nova rather than a PCEB \citep[][ see also
Sect.\,4]{kawkaetal02-1}.

As expected, $\tsd$ of an individual system depends strongly on
the assumed AML prescription.  Following \citet{andronovetal03-1} we
obtain shorter evolution times (by a factor of $1.5-4$) 
compared to the standard
scenario for systems with $\Msec<0.3\Msun$, whereas for systems with
$\Msec\geq0.3\,\Msun$
\tsd\ is significantly longer (by a factor of $3-30$ depending on the
present orbital period of the system). The differences in \tsd,
depending on the use of \Jcmb\ or \Jrmb\  are illustrated for
the systems HR\,Cam and EG\,UMa in Fig.\,\ref{f-amlcmp}. 

Figure\,\ref{f-teffcon} shows that most of the PCEB white dwarfs
will be very cool when the secondary is expected to fill its
Roche-lobe and mass transfer starts. Only for \Jcmb we get one system,
(i.e. V471\,Tau, the filled square in the upper left corner of
Fig.\,\ref{f-teffcon}) in which the mass transfer starts while the
primary still can be considered a ``hot'' white dwarf.

\begin{figure}
\includegraphics[angle=0,width=8.7cm]{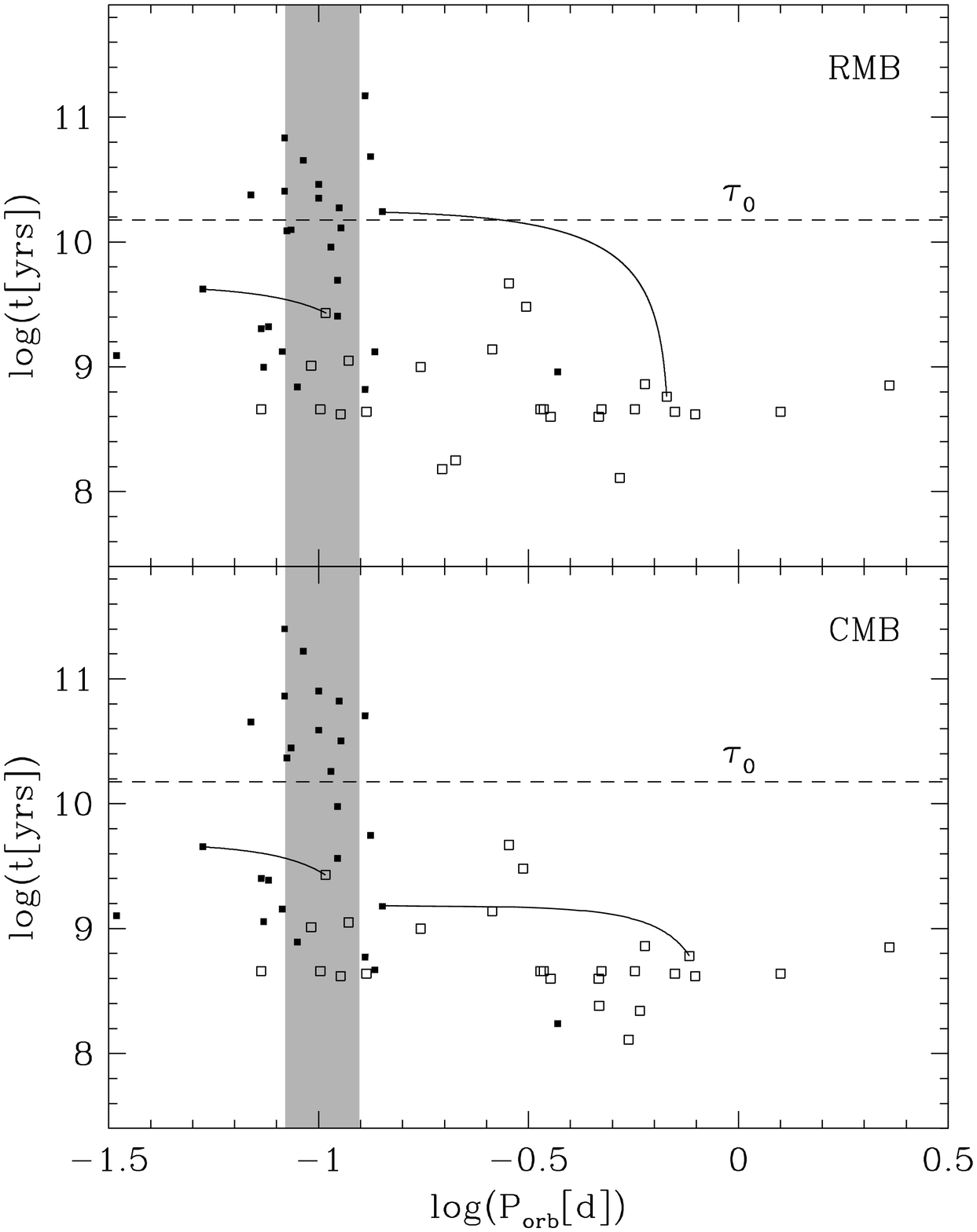}
\caption{\label{f-pce_psd} Age of the binary versus orbital period for
\Jcmb\ (bottom panel) and \Jrmb\ (top panel). The open squares mark the
positions of the PCEBs immediately after the CE-phase
($P_{\mathrm{CE}}$, $t_{\mathrm{evol}}$) whereas the filled squares
indicate the positions when the mass transfer starts ($\Psd$,
$t_{\mathrm{evol}}+t_{\mathrm{cool}}+\tsd$). The solid lines
show the orbital evolution of EG\,UMa and HR\,Cam. The dashed
horizontal line indicates the age of the universe
$\tau_0\sim\,1.3\times10^{10}$\,yrs (see text). }
\end{figure}

\begin{table*}
\caption{\label{t-future} The properties of the PCEBs calculated in
Sect.\,5--7. If there is no reference given in the Table, the distance
is obtained from the references listed in Table\,1. The time until the
mass transfer is expected to start ($\tsd$) is calculated using
Eq.\,(\ref{eq-sol1}), (\ref{eq-sol2}) (CMB) and Eq.\,(\ref{eq-sol3})
(RMB).  The total time until the system becomes semi-detached,
$t_{\mathrm{tot}}$ is given by the sum of the present cooling age, 
$t_{\mathrm{cool}}$, the estimated MS lifetime of the primary, 
$t_{\mathrm{evol}}$,  and $\tsd$. PCEBs with $t_{\mathrm{tot}}$ less than a 
Hubble time are marked in bold.
For the systems RR\,Cae, EG\,UMa, EC\,13471--1258, BPM\,71214 and
V471\,Tau we find the properties of the progenitor of the PCEB
(orbital period $P_{\mathrm{CE}}$, primary mass $M_{1,\mathrm{pCE}}$
and, $t_{\mathrm{evol}}$) depending on the assumed AML (see text).  In
these cases, the second line gives the results assuming reduced
magnetic braking (RMB) whereas the first line corresponds to the
classical magnetic braking assumption (CMB). 
$f_{\mathrm{PCEB}}$ gives the mean fractional PCEB
life time that the system has already passed through.
The orbital period at
which the PCEB starts mass transfer $\Psd$ is independent on the
assumed angular momentum loss.}
\newcommand{\pc}[1]{\textbf{#1}}
\setlength{\tabcolsep}{0.69ex}
\begin{tabular}{llllrrrrrrrrrl}
\hline\noalign{\smallskip}
&&&&&&&&
\multicolumn{2}{l}{\large{CMB}} &&
\multicolumn{2}{l}{\large{RMB}} &
\\
Object&
\multicolumn{1}{l}{$d$[pc]} &
Ref. &
\multicolumn{1}{c}{$\log(t_{\mathrm{cool}})$} &
\multicolumn{1}{c}{$P_{\mathrm{CE}}$[d]} &
\multicolumn{1}{c}{$\log(t_{\mathrm{evol}})$} &
\multicolumn{1}{c}{$M_{1,\mathrm{pCE}}$} &
\multicolumn{1}{c}{$\log(\tsd)$} &
\multicolumn{1}{c}{$f_{\mathrm{PCEB}}$} &
\multicolumn{1}{c}{$\log(t_{\mathrm{tot}})$} &
\multicolumn{1}{c}{$\log(\tsd)$} &
\multicolumn{1}{c}{$f_{\mathrm{PCEB}}$} &
\multicolumn{1}{c}{$\log(t_{\mathrm{tot}})$} &
\multicolumn{1}{c}{$\Psd$[d]} \\
\noalign{\smallskip}\hline\noalign{\smallskip}
RR\,Cae  & 11          & 1   & 9.07 & 0.307 & 9.48 & 1.41 & 10.54      & 0.033
& 10.58& 10.26      & 0.061 & 10.35    & 0.100 \\
         &             &     &      & 0.312 & 9.49 & 1.41 &            &
&      &            &       &          &  \\
EG\,UMa  & $32\pm5$ & 2   & 8.52 & 0.764 & 8.78 & 2.58 & 8.76  & 0.365
& \pc{9.18}  & 10.22      & 0.020 & 10.24   & 0.142 \\
&&&& 0.675 & 8.76 & 2.63 &&&&&&&\\
EC 13471--1258 & 55 &               & 8.57 & 0.583 & 8.34 & 3.98 & 
6.23 & 0.995 & \pc{8.77} &  8.16 & 0.720 & \pc{8.82} & 0.129 \\
&&&& 0.197 & 8.18 & 4.71 &&&&&&&\\
BPM\,71214 & 68 &               & 8.33 & 0.466 & 8.38 & 3.79 & 
7.00 & 0.955 & \pc{8.67} &  8.97 & 0.186 & \pc{9.12} & 0.136 \\    
&&&& 0.212 & 8.25 & 4.37 &&&&&&&\\
HR\,Cam  & $72\pm11$     & 2,3 & 7.65 & 0.104 & 9.43 & 1.48 & 9.25  & 0.025
& \pc{9.66}  & 9.16  & 0.030 & \pc{9.62}   & 0.053 \\
UZ\,Sex  & $35\pm5$      & 2   & 7.92 & 0.599 & 8.86 & 2.39 & 10.90      & 0.001
& 10.90 & 10.45      & 0.003 & 10.46   & 0.100 \\
BPM\,6502& $25\pm4$      & 2   & 7.68 & 0.338 & 8.66 & 2.88 & 10.44      & 0.002
& 10.45  & 10.08 & 0.004 & \pc{10.10}   & 0.086 \\
HZ\,9    & $40\pm6$      & 2   & 8.02 & 0.567 & 8.66 & 2.88 & 10.82      & 0.002
& 10.82  & 10.26      & 0.006  & 10.27  & 0.112 \\
         & 46            & 3   &      &       &      &      &            &
&               &          &      & & \\
MS\,Peg  & 61          &     & 7.45 & 0.175 & 9.00 & 2.12 & 9.42  & 0.011
& \pc{9.56}  & 9.18  & 0.018  & \pc{9.40}  & 0.111 \\
CC\,Cet  & $89\pm13$   & 2   & 7.05 & 0.284 & 9.67 & 1.22 & 10.27      & 0.001
& 10.36  & 9.88  & 0.001 & \pc{10.09}  & 0.084 \\
HW\,Vir  & $171\pm19$  &     &      & 0.118 & 9.05 & 2.02 & 9.12  &
& \pc{9.38}  & 8.99  &    & \pc{9.32}      & 0.076 \\
HS\,0705+6700 & 1150   & 2   &      & 0.096 & 9.01 & 2.11 & 8.61  &
& \pc{9.16} & 8.48  &      & \pc{9.12}    & 0.082 \\
LM\,Com  & $170\pm26$  & 2   & 7.05 & 0.259 & 9.14 & 1.87 & 9.91  & 0.001
& \pc{9.98}  & 9.55  & 0.003 & \pc{9.69}   & 0.111 \\
         & 290         &     &  &&&&&&&&&&\\
PG\,1017--086 &990     & 2   &      & 0.073 & 8.66 & 2.88 & 8.91  &
 & \pc{9.10} & 8.89  &      & \pc{9.09}    & 0.033 \\
V471\,Tau &$47\pm4$    & 3   & 6.93 & 0.584 & 8.11 & 5.08 & 7.54 & 0.197
& \pc{8.23}   & 8.89  & 0.011  & \pc{8.91}  & 0.371 \\
&&&& 0.522 & 8.10 & 5.12 &&&&&&&\\
NY Vir   &$710\pm50$   &     &      & 0.101 & 8.66 & 2.88 & 8.51  &
& \pc{8.89}  & 8.37  &    & \pc{8.84}      & 0.089 \\
         & 560         & 2   &  &&    & & & &            &          &            &          &       \\
AA\,Dor  & 396         &     &    &&   & & 10.65      &  &        & 10.51  &    &          & 0.046 \\
RE\,2013+400& 95--125   &     & 7.08 & 0.706 & 8.64 & 2.95 & 11.22      &
$<0.001$ & 11.22 & 10.65      & $<0.001$ & 10.65 & 0.092 \\
GK\,Vir  & $350\pm50$  & 2   & 6.20 & 0.344 & 8.66 & 2.89 & 10.65      &
$<0.001$ & 10.65 & 10.37      & $<0.001$ & 10.38 & 0.069 \\
MT\,Ser  & 4300        & 4   &      & 0.113 & 8.62 & 3.00 & 8.86  &
& \pc{9.06}  & 8.76  &      & \pc{9.00}    & 0.074 \\
         & 4510        & 5   &    &&  &   & & &          &          &            &          &       \\
         & 5400        & 6   &     && &   & & &         &          &            &          &       \\
         & 4600        & 7   &      &&&   & & &         &          &            &          &       \\
IN\,CMa  & 158-208     &     & 6.10 & 1.260 & 8.64 & 2.96 & 9.71  &
$<0.001$ & \pc{9.75} & 10.68      & $<0.001$ & 10.68 & 0.133 \\
         & 186         & 8   &      &    & & &  &&       &          &
&          &       \\
NN\,Ser  & 356--417     & 9  & 6.12 & 0.130 & 8.64 & 2.96 & 9.32   & 0.001
& \pc{9.40}  & 9.20  & 0.001  & \pc{9.31}  & 0.073 \\ 
TW\,Crv  & 557         &     &      &&&   & & &          &          &            &          &       \\
RE\,1016--053& 75--109  &     & 6.19 & 0.790 & 8.62 & 3.00 & 11.40      &
$<0.001$ & 11.40 & 10.83      & $<0.001$ & 10.83 & 0.083 \\
UU\,Sge  & $2400\pm400$&     &      & 0.465 & 8.60 & 3.05 & 10.50      &
& 10.51 & 10.10      &      & \pc{10.11}     & 0.113 \\
         & 3600        & 10  &      &      &&&  &&    &          &            &          &       \\
         & 2150        & 5   &      &      &&&   &&   &          &            &          &       \\
V477\,Lyr &$1700\pm600$&     &      & 0.472 & 8.66 & 2.89 & 10.86      &
& 10.86 & 10.40      &      & 10.41    & 0.083 \\
          & 1850       & 5   &     && &      &&&      &          &            &          &       \\
          & 2190       & 8   &      &&&      &&&      &          &            &          &       \\
PN A66 65 & 1100       & 5   &      &  &&    &&&      &          &            &          &       \\
          & 1660       & 8   &      &    &&  &&&      &          &            &          &       \\
KV\,Vel   & 730-1000   & 11  &      & 0.357 & 8.60 & 3.05 & 10.25      &
& 10.25   & 9.94  &  & \pc{9.95}  & 0.107 \\
HS\,1136+6646 &        &     & 5.55 &   &&   &&&      &          &            &
     &       \\
BE\,UMa   & 2000       &     &      & 2.291 & 8.85 & 2.42 & 10.70      &
& 10.71  & 11.17      &   & 11.71   & 0.129 \\
\noalign{\smallskip}\hline\noalign{\smallskip}
\end{tabular}
\linebreak References: (1) \citet{bruch+diaz98-1}, (2) this work,
section\,\ref{s-distance}, (3) \citet{perrymanetal98-1}, (4)
\citet{abell66-1}, (5) \citet{cahn+kaler71-1}, (6) \citet{maciel84-1},
(7) \citet{cahnetal92-1}, (8) \citet{vennes+thorstensen96-1}, (9)
\citet{wood+marsh91-1}, (10) \citet{waltonetal93-1}, (11)
\citet{pottasch96-1}.
\end{table*}
 
\subsection{Representative progenitors of the current CV population}

Analysing PCEBs is not only interesting in itself, it is also
important in the context of CV formation and evolution (see Sect.\,1
and \,2).  Discussing the properties of our sample of PCEBs in the
framework of the {\em present--day} CV population requires to select
those systems which are representative for the former progenitors of
the current CV population.  Only PCEBs which will evolve into CVs in
less than the Hubble time (assuming
$\tau_0=1.3\times10^{10}$\,yrs\footnote{Recently
\citet{ferrerasetal01-1} set new constraints on the age of the
universe combining the results obtained by cosmochromology, stellar
population synthesis, and mapping of the peaks in the microwave
background.  They derived an age of $13.2^{+1.2}_{-0.8}$\,Gyrs.  Note
also that \citet{cayreletal01-1} obtained a stellar age of
$12.5\pm3$\,Gyrs from uranium decay. Throughout this work we therefore
assume that star formation in the Galaxy began $13$\,Gyrs ago.}) 
satisfy this condition. To distinguish these systems from the PCEBs
which are different from the former progenitors of the present CVs we
refer to them as \textit{pre-CVs} (see also Sect.\,1).

Applying this selection criterion, we find 14 (16) pre-CV candidates in
our sample (Table\,\ref{t-future}) when assuming \Jcmb\
(\Jrmb). Although \tsd\ depends strongly on the assumed AML for some
individual systems (e.g. EG\,UMa, Fig.\,\ref{f-amlcmp}; see also
Fig.\,\ref{f-pce_psd}), the number of pre-CVs among the PCEB sample is
nearly independent on it (see Table\,\ref{t-future},
Fig.\,\ref{f-pce_psd}). From the sample of pre-CVs 5 (3) will
initiate mass transfer at orbital periods $>3$\,h whereas the other 9
(13) systems will start their CV life in or below the orbital period
gap.  This finding is in good agreement with the prediction of
\citet{kingetal94-1} that $\gappr67\%$ of CVs will start mass transfer
near or below the period gap. Possible observational selection effects
will be discussed in Sect.\,9.

A further test on whether the current pre-CV population is a
representative sample of progenitors for the present CV population
comes from comparing their space density to that of CVs. An estimate
of the space density of PCEBs/pre-CVs requires obviously the knowledge
of their distances.

\section{\label{s-distance}Distance estimates}

We estimate the distances to the systems in Table\,1 for
which no previous distance determination has been available in the
literature using the following methods. 

\subsection{PCEBs with a white dwarf primary}
For all but one system without a published distance estimate
ultraviolet (UV) spectroscopy is available in the public archives of
the \textit{International Ultraviolet Explorer} (\textit{IUE}) and the
\textit{Hubble Space Telescope}. We fitted the UV data listed by
Table\,3 with pure-hydrogen white dwarf model spectra computed with
the code described by \citet{gaensickeetal95-1}, fixing $\log g=8.0$,
i.e. $\Mwd\approx0.6\,\Msun$. The free parameters of this fit were,
hence, the white dwarf temperature \Teff\ and the scaling factor
$f/H=4\pi\Rwd^2/d^2$, with $f$ the observed flux and $H$ the Eddington
flux of the model spectra. The results from our fits are given in
Table\,3. Overall, the white dwarf temperatures that we derived here
are in good agreement with the published values.

\begin{figure}
\includegraphics[angle=-90,width=8.8cm]{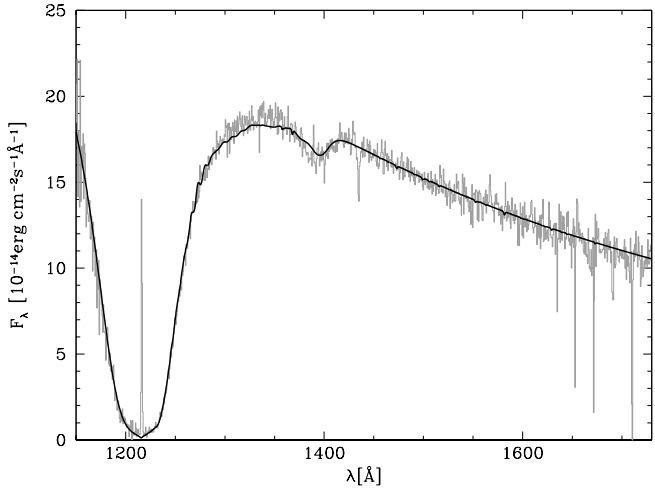}
\caption{\label{f-hz9} The HST/STIS echelle spectrum of HZ9 along with the best-fit
$\Teff=17\,400$\,K pure hydrogen white dwarf model, assuming $\log
g=8.0$. The sharp ``absorption dips'' at the red end of the spectrum
are due to gaps between the echelle orders.}
\end{figure}

Fixing the surface gravity (\,=\,white dwarf mass) in the fit
introduces a systematic uncertainty in the derived distances, as the
white dwarf radius obviously depends on the assumed mass. However, it
is important to notice that also the white dwarf temperature derived
from the fit depends on the assumed mass, with higher (lower)
temperatures resulting for higher (lower) $\log g$. This effect
compensates to some extent the $\Rwd(\Mwd)$ dependence, and the error
in the spectroscopic parallaxes is $\simeq\pm15\%$ for an assumed
range $\log g=8.0\pm0.5$ (i.e. $0.35\,\Msun\la\Mwd\la0.9\,\Msun$),
which covers probably most of the analysed objects.
The Hyades member HZ\,9 can be used to estimate the robustness of our
distance estimate and error analysis. From our model fit to
unpublished \textit{HST} STIS spectroscopy, we find a temperature
which is slightly below the low end of the range quoted by
\citet{guinan+sion84-1}, which was based on the analysis of
\textit{IUE} spectroscopy. However, the much better \textit{HST} data
available now clearly shows the 1400\,\AA\ quasimolecular $H_2+$
absorption which indicates a temperature $<20\,000$\,K
(Fig.\,\ref{f-hz9}). We derive from the scaling factor of the model a
distance of $d=40\pm6$\,pc, which is entirely consistent with the
Hipparcos-measured parallax \citep{perrymanetal98-1}.

For GK\,Vir, we had to fix the temperature to the value derived by
\citet{fulbrightetal93-1} as the temperature-sensitive \La\ absorption
line is entirely blended with geocoronal emission in the available \textit{IUE}
spectrum.  

A final note concerns the distance of LM\,Com. \citet{oroszetal99-1}
quote a distance of 290--308\,pc, which they base on a lengthy
discussion of the white dwarf contribution to the observed $R$ band
magnitude. However, as the contamination of the M-dwarf is lowest in
the blue, we re-examined the distance of LM\,Com scaling a
$\Teff=30\,000$\,K, $\log g=8.0$ model to
$F_{\lambda}(4700\AA)\simeq17\times10^{-16}$\,\ecsa\ (measured from
\citealt{oroszetal99-1} Fig.\,7), and find $d=170$\,pc.

\subsection{PCEBs containing a sdB primary} 
Considering the rather similar properties of the sdB primaries in
HW\,Vir, HS\,0705+6700, PG\,1017--086, and NY\,Vir (\Mwd, \Twd, see
Table\,1), we computed (crude) distances estimates by scaling the
distance of HW\,Vir ($d=171$\,pc, $V=10.6$; \citealt{wood+saffer99-1})
to the observed $V$ magnitudes of the other sdB PCEBs (HS\,0705+6700:
$V=14.8$; PG\,1017--086: $V=14.4$; NY\,Vir: $V=13.2$). This estimate
ignores possible extinction as well as differences in $\Teff$ and
$\Rwd$.

\begin{table}
\caption[]{Distance estimates for PCEBs, see text}
\setlength{\tabcolsep}{1.9ex}
\begin{tabular}{lrrrr} 
\noalign{\smallskip}\hline\noalign{\smallskip}
Object   & Instrument & Dataset   & \Teff\,[K] & d\,[pc] \\
\noalign{\smallskip}\hline\noalign{\smallskip}
EG UMa   & HST/FOS    & y16u0502t & 13\,400      & 32 \\
HR Cam   & HST/STIS   & o6gj02020 & 20\,800      & 72 \\
UZ Sex   & IUE        & swp27393  & 17\,200      & 35 \\
BPM 6502 & IUE        & swp27351  & 21\,400      & 25 \\
HZ 9     & HST/STIS   & o5dma6010 & 17\,400      & 40 \\
CC Cet   & IUE        & swp27392  & 25\,000      & 89 \\
GK Vir	 & IUE        & swp07459  & 48\,800$^{*}$      & 350\\
\noalign{\smallskip}\hline\noalign{\smallskip}
\end{tabular}
$^{*}$Temperature fixed to the value given by
\citet{fulbrightetal93-1}.
\end{table}

\section{\label{s-space_densities}Space densities}
Inspection of Table\,\ref{t-future} shows that 12 PCEBs from our
sample are within 100\,pc (including RE\,1016-053, $d=75-109$\,pc).
This gives an ``observed'' space density of
($=2.9\times10^{-6}\mathrm{pc^{-3}}$) which can be considered to be a
conservative lower limit on the PCEB space density as
the presently known PCEB sample is systematically dominated by young
PCEBs (see Sect.\,5.1, Table\,\ref{t-future}).  Accounting for this
bias, we estimate the true PCEB space density to be:
\begin{equation}\label{eq-space1}
\rho_{\mathrm{PCEB}}\simeq2.9\times10^{-6}\frac{0.5}{\overline{f}_{\mathrm{PCEB}}}=
\left\{
\begin{array}{rr}
6.6\times10^{-6}\mathrm{pc^{-3}} \hspace{0.2cm}(\Jcmb)\\
2.9\times10^{-5}\mathrm{pc^{-3}} \hspace{0.2cm}(\Jrmb)
\end{array}\right.
\end{equation}
where $\overline{f}_{\mathrm{PCEB}}$ is the mean fractional PCEB
life time that the systems have already passed through.
%
%
The substantial difference between the estimates of
$\rho_{\mathrm{PCEB}}$ for the two AML prescriptions results from the
fact that the mean evolution time scale is significantly longer for
$\dot{J}=\Jrmb$ 
and, thus, the existence of more old and yet undetected PCEBs is 
predicted.  However,
even the higher value of $\rho_{\mathrm{PCEB}}$, assuming \Jrmb, is
still significantly below the theoretical predictions of \citet[][
their Fig.\,4]{dekool+ritter93-1} indicating that our sample is
probably not only biased towards young systems (see
Sect.\,\ref{s-finding_pcebs}).

Requiring $t_{\mathrm{tot}}<\tau_0$ for PCEBs which can be considered
pre-CVs, we find 6 (\Jcmb) and 7 (\Jrmb) of these systems within 100\,pc,
resulting in lower limits on their space density
$\rho_{\mathrm{preCV}}>1.4\times10^{-6}\mathrm{pc^{-3}}$ (\Jcmb) and
$\rho_{\mathrm{preCV}}>1.7\times10^{-6}\mathrm{pc^{-3}}$ (\Jrmb). 
We estimate the actual pre-CV space density, again taking into account
age and evolution time scale and find 
\begin{equation}
\rho_{\mathrm{preCV}}\simeq
\left\{\begin{array}{ll}
1.7\times10^{-6}\mathrm{pc}^{-3} \hspace{0.4cm}(\Jcmb)\\
6.0\times10^{-6}\mathrm{pc}^{-3} \hspace{0.4cm}(\Jrmb).
\end{array}\right.
\end{equation}

Our estimate of the current pre-CV space density allows an 
estimate for the present CV space density. Considering the average of
the pre-CV lifetime ($\overline{t}_{\mathrm{preCV}}$) of the systems
we find within 100\,pc and following \citet{politano96-1} in assuming
that the CV birthrate has been constant since
$t_{\mathrm{gal}}=10^{10}$\,yrs we get
\begin{equation}
\rho_{\mathrm{CV}}\sim\,t_{\mathrm{gal}}\frac{\rho_{\mathrm{preCV}}}{\overline{t}_{\mathrm{preCV}}}\sim\,1\times10^{-5}\,\mathrm{pc}^{-3}  
\end{equation} 
for both angular momentum loss prescriptions.
 
Considering the assumptions involved and the small number statistics
of the known pre-CV sample, this result has to be regarded as a rather
rough estimate. Nevertheless, our estimate for $\rho_{\mathrm{CV}}$ is
in agreement with the high end of current observational estimates
\citep[see][]{gaensickeetal02-2} but below theoretical predictions:
$10^{-4}\mathrm{pc}^{-3}$ \citep{dekool92-1} to
$2\times10^{-5}\mathrm{pc}^{-3}$ \citep{politano96-1}.  As we will
discuss in Sect.\,\ref{s-lackofearlysecs}, the presently known sample
of PCEBs/pre-CVs is very likely incomplete not only with respect to
those systems containing cold white dwarfs, but also to those systems
containing an early-type secondary. Hence, we would like to stress
that all the space densities derived above should be regarded as lower
limits to the true values.

Considering the binary age postulate (BAP) of
\citet{king+schenker02-1} outlined in Sect.\,2 we note that assuming
reduced magnetic braking indeed leads to an increase of the averaged
evolution time scale for pre-CVs but the presently known PCEB population is too
small and too strongly biased to prove or disprove the BAP scenario.

\begin{figure*}
\begin{center}
\includegraphics[width=7.5cm, angle=270]{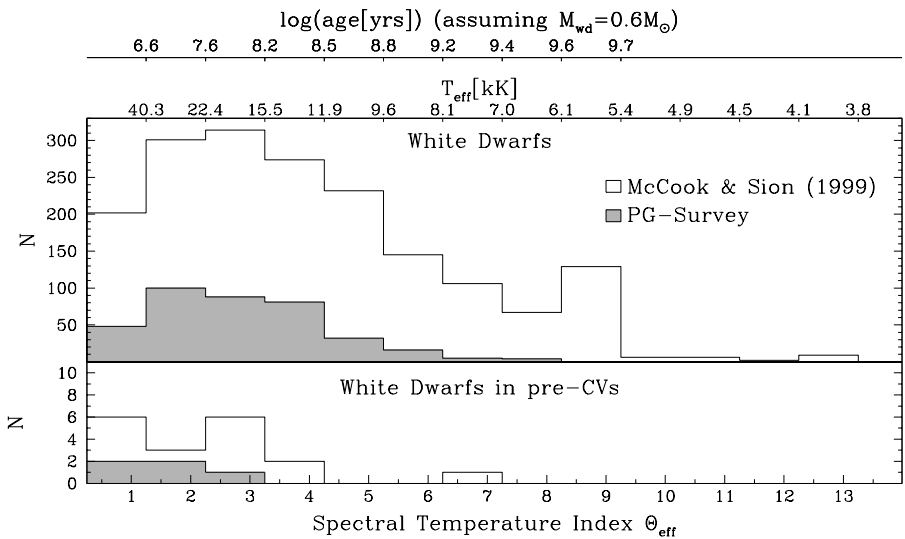}
\caption{\label{f-wd_pceb_teff} The currently known populations of single
white dwarfs (top) and PCEBs (bottom) as a function of their effective
temperature. Here, we follow \citet{mccook+sion99-1} and plot the
spectral temperature index, $\Theta_{\mathrm{eff}}\,=50\,400/\Teff$,
which is a direct indicator of the age of the white dwarfs.  The grey
shaded histograms represent systems which have been found in the PG
survey. Obviously, there exists a population of cool single white
dwarfs whereas PCEBs white dwarfs are generally hot and young.}
\end{center}
\end{figure*}

\section{\label{s-selection}Observational selection effects}

We have shown that the known population of PCEBs is dominated by young
hot systems (Sect.\,\ref{s-coolingage}), and that, applying the
current theories of (pre-)CV evolution, the majority of them will
remain in a detached configuration for many $10^8$\,yrs to several
$10^9$\,yrs (Sect.\,\ref{s-future})~--~sufficiently long for their
white dwarf primaries to cool --~depending on \tsd\ and \Mwd~-- to
temperatures $6000\,\mathrm{K}\la\Teff\la15\,000$\,K (see
Fig.\,\ref{f-teffcon})
These findings strongly suggest that~--~if the reality about AML in
PCEBs is somewhere bracketed between the prescriptions of \Jcmb\
and \Jrmb~--~the currently known population of PCEBs is highly
incomplete.

As a test for possible selection effect, we compare the effective
temperature distribution of the white dwarfs in our PCEB sample to
that of field white dwarfs.

\subsection{\label{s-pceb_vs_ms}
The known PCEB population vs. the \citet{mccook+sion99-1}
white dwarf population} 
\citet{mccook+sion99-1} present a catalogue of 1793 spectroscopically
identified white dwarfs, including information on the white dwarf
temperatures where available. The temperature distribution of the
\citeauthor{mccook+sion99-1} white dwarf sample peaks at temperatures
$T_{\mathrm{eff~WD}}\simeq\,15\,000-22\,000$\,K (top panel of
Fig.\,\ref{f-wd_pceb_teff}) with a mean effective temperature of
$\overline{T}_{\mathrm{eff~WD}}=19\,360$\,K. Also shown in
Fig.\,\ref{f-wd_pceb_teff} (bottom panel) is the temperature
distribution of the PCEB white dwarf primaries from our sample
(Table\,\ref{t-sample}), 
which also peaks in the range
$T_{\mathrm{eff~PCEB}}\simeq\,15\,000-22\,000$\,K, but lacks the
low-temperature tail seen in the \citeauthor{mccook+sion99-1} sample.
The average temperature of the white dwarfs in our PCEB sample is 
$\overline{T}_{\mathrm{eff~PCEB}}=33\,423$\,K. 
Using $\chi^2$ statistics, the significance that two distributions are
different is
\begin{equation}
P\left(\frac{\nu}{2},\frac{\chi^2}{2}\right)=\frac{\int_0^{\chi^2/2}e^{-t}t^{\nu/2-1}dt}{\int_0^{\infty}t^{\nu/2-1}e^{-t}dt},
\end{equation}
where $\nu$ is the number of degrees of freedom, i.e. the number of bins, and
$\chi^2$ is given by:
\begin{equation}
\chi^2=\sum_{i=1}^{\nu}\frac{\sqrt{N_{\mathrm{pc}}/N_{\mathrm{wd}}}N_{\mathrm{wd,}i}-\sqrt{N_{\mathrm{wd}}/N_{\mathrm{pc}}}N_{\mathrm{pc,}i}}{N_{\mathrm{pc},i}+N_{\mathrm{wd,}i}},
\end{equation}
with $N_i$ the number of objects in bin $i$,
$N_{\mathrm{wd}}\equiv\,\sum_{i=1}^{\nu}N_{\mathrm{wd,i}}$ and
$N_{\mathrm{pc}}\equiv\,\sum_{i=1}^{\nu}N_{\mathrm{pc,i}}$.  We obtain
for the probability that the two distributions are different:
\begin{equation}
P\left(\frac{\nu}{2},\frac{\chi^2}{2}\right)=0.93.
\end{equation}

It appears, hence, likely that the PCEB white dwarfs are indeed on
average hotter --~and therefore younger~-- than the single field white
dwarfs of \citep{mccook+sion99-1}.

\subsection{\label{s-pceb_vs_pg}
PCEB and field white dwarfs in the Palomar Green survey}

Whereas the \citet{mccook+sion99-1} is a heterogenous catalogue of
white dwarfs identified by various means, it is of fundamental
interest to compare the properties of white dwarfs in PCEBs and of
field white dwarfs drawn from a single survey. Inspection of
Table\,\ref{t-sample} shows that five of our PCEBs with a white dwarf
primary are contained in the Palomar-Green (PG) survey
\citep{greenetal86-1}: UZ\,Sex, CC\,Cet, LM\,Com, GK\,Vir and
NN\,Ser. Fig.\,\ref{f-wd_pceb_teff} shows the temperature distribution
of both the field white dwarfs (upper panel) and of the PCEB white
dwarfs (bottom panel) from the PG survey.  The mean values of the
effective temperatures for PCEB white dwarfs and single white dwarfs
contained in the PG survey are
$\overline{T}_{\mathrm{eff~PCEB}}=35\,380$\,K and
$\overline{T}_{\mathrm{eff~WD}}\sim\,21\,000$\,K, respectively. 
The mean effective temperature of the white dwarfs in PCEBs 
is again somewhat higher than that of their field relatives.
As there are only five PCEBs with a white dwarf primary 
included in the PG-survey, the statistic is very poor but for completeness 
we give the probability that the difference of the two distributions (see the
shaded histograms in Fig.\,\ref{f-wd_pceb_teff}) can be explained by
chance
\begin{equation}
1-P\left(\frac{\nu}{2},\frac{\chi^2}{2}\right)=0.51.
\end{equation}
Obviously, from this subset of systems we can neither say 
that the white dwarfs in PCEBs are 
{\em systematically} hotter and younger than their field relatives nor can 
we exclude that this is the case. 
In the next section we will discuss observational selection effects.
This will give us a hint which possibility is more likely.

Notice, the probabilities given by Eq.\,(17) and (18) are calculated by
neglecting the very few detected extremely old single white dwarfs,
i.e. using only 9 (Eq.\,(17)) respectively 5 (Eq.\,(18)) bins.

\begin{figure*}
\includegraphics[angle=270,width=9cm]{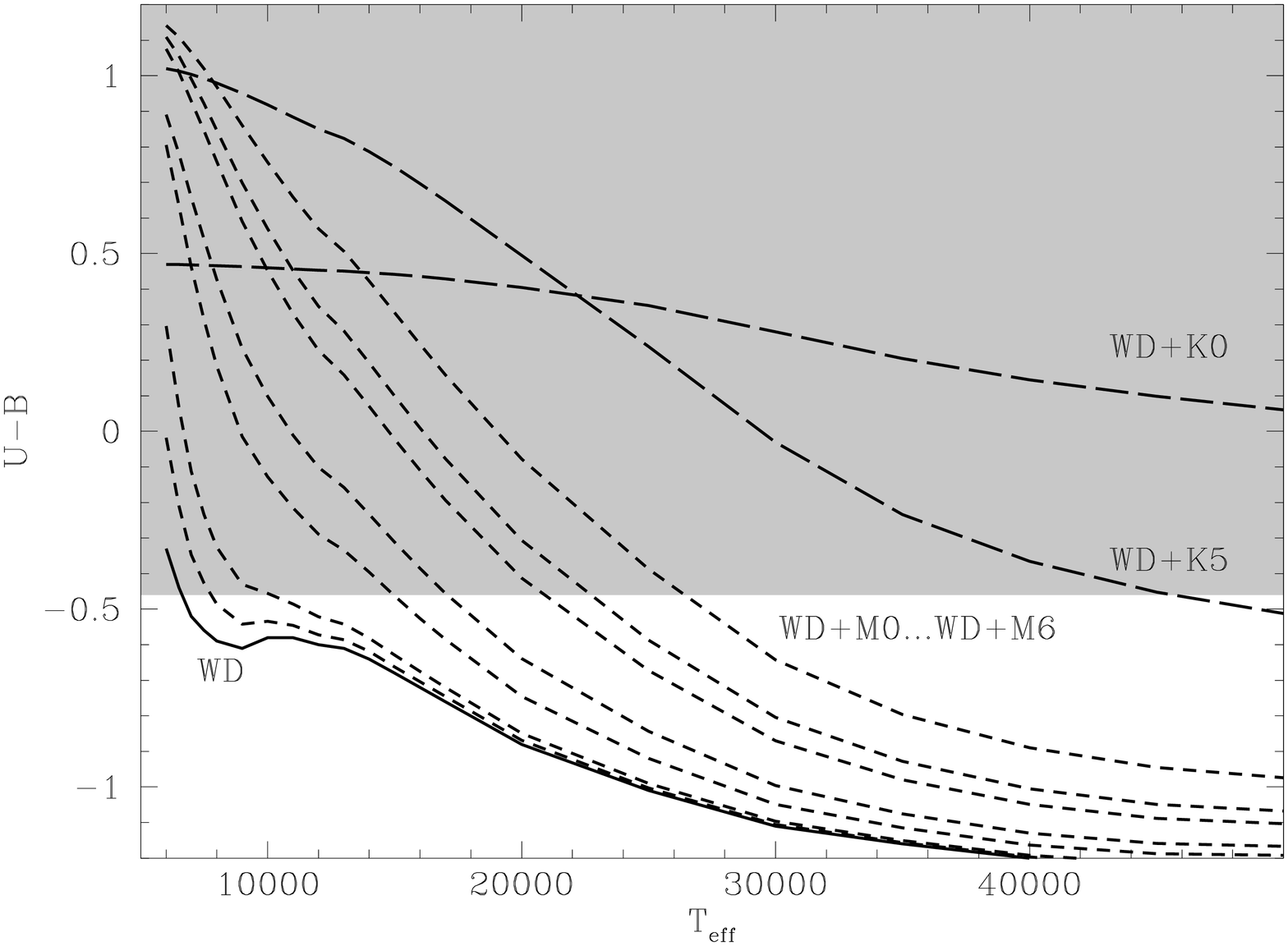}
\hfill
\includegraphics[angle=270,width=9cm]{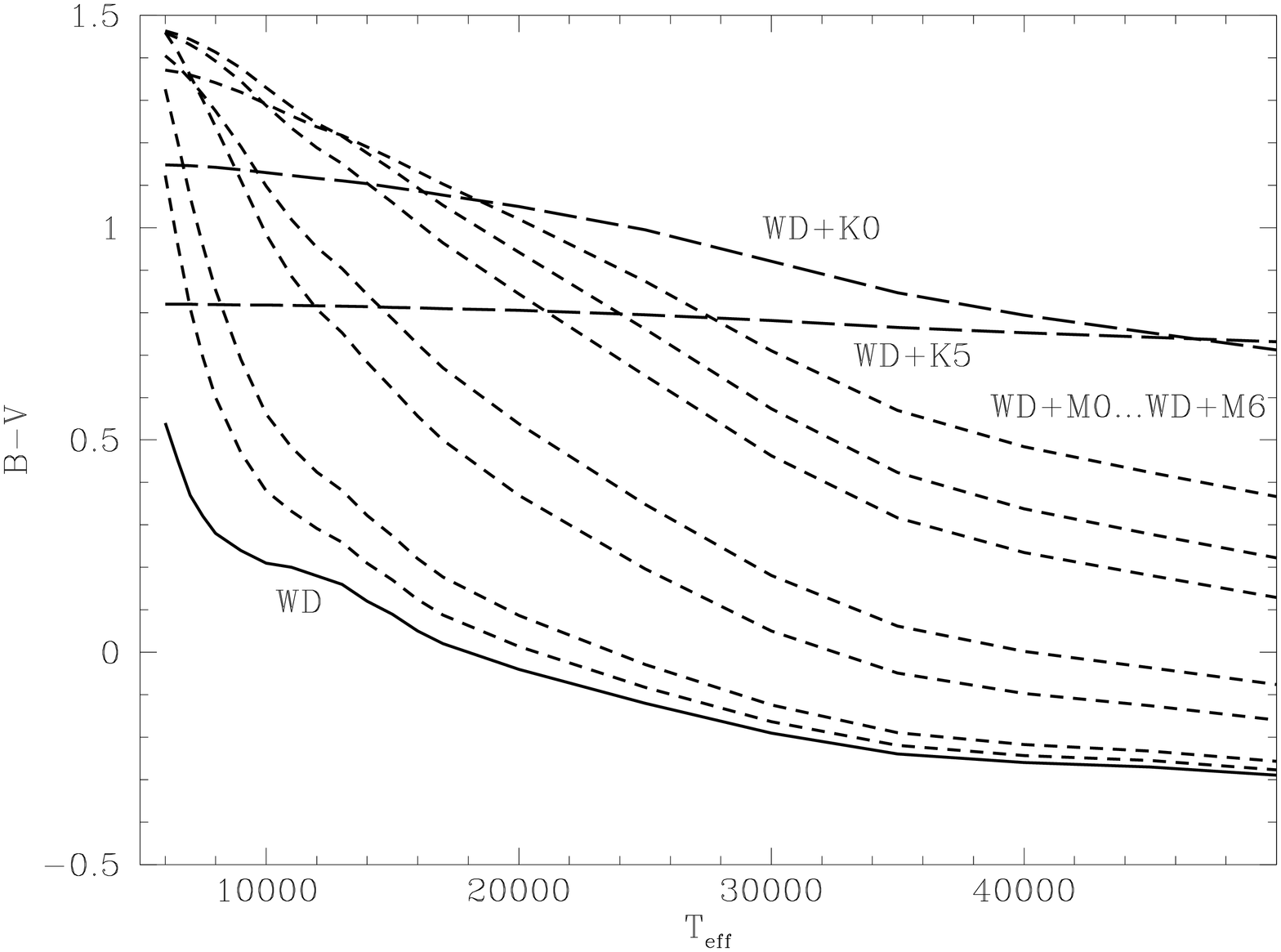}
\caption{\label{f-pceb_colours} Simulated colours for PCEBs containing
a white dwarf with $6000\,\mathrm{K}<\Teff<50\,000$ and a companion with a
spectral type in the range K0 to M6. The left panel shows in gray the
PCEB parameter space that \textit{do not} fulfill the selection
criterion of the PG survey ($U-B<0.46$). }
\end{figure*}

\subsection{\label{s-finding_pcebs}Finding PCEBs}
Based on the presently available data (Sect.\,\ref{s-pceb_vs_ms} and
\ref{s-pceb_vs_pg}), we can not exclude that the white dwarfs in PCEBs
are --~on average~-- \textit{systematically hotter} (and younger) than
single white dwarfs. This statement may be entirely based on selection
effects skewing the temperature/age distribution of the white dwarfs
in the known PCEB population. It is, however, very difficult to assess
this bias quantitatively.  We list in Appendix\,A for the PCEBs from our
sample in Table\,\ref{t-sample} (1) the way that they were discovered
in the first place, and (2) the way that their binarity has been
identified. Inspection of these notes shows that the vast majority of the
known PCEBs have been identified in the first place \textit{because}
of their white dwarf component~--~either as blue object or X-ray
source.  Obviously, the presence of a late-type companion may
significantly affect the colour of a PCEB: the cooler the white dwarf
and/or the earlier the companion are, the less \textit{blue} the system
will appear. This may affect different candidate samples in different
ways, depending on how the criterion \textit{blue} has been defined.

We have computed the colours expected for PCEBs containing a white
dwarf with $6000\,$K$<\Teff<50\,000$\,K and an ZAMS companion with
spectral type K0 to M6, the results for $U-B$ and $B-V$ are shown in
Fig.\,\ref{f-pceb_colours}. Comparing these simulated colours to the
selection criterion of the PG survey, $U-B<0.46$
\citep{greenetal86-1}, it becomes apparent that PCEBs containing a
cool ($\Teff\la15\,000$\,K) white dwarf will be included in the PG
survey only if their companions are of spectral type $\sim$M4 or
later. Similarly, for such PCEBs $B-V>0.5$, which will most likely
prevent their selection as white dwarf candidates.  As an example for
the ``historic'' white dwarf searches, \cite{giclasetal65-1} used a
Lowell colour class $-1$ or 0 (corresponding to $B-V<0.0$ or
$B-V<+0.2$, and $U-B<-0.78$ or $U-B<-0.60$) as selection
criterion. Consequently, the GD lists will contain only those PCEBs
containing moderately hot white dwarfs and late type secondaries, such
as HR\,Cam and MS\,Peg.

PCEBs which were not initially selected as white dwarf candidates are:
(1) the planetary nebulae; (2) high proper motion objects (RR\,Cae,
BPM\,6502, and BPM\,71214)~--~interestingly enough, RR\,Cae is the
PCEB containing the coldest white dwarf in our sample; and (3) variable
stars (V471\,Tau: spectroscopic; BE\,UMa: photometric). In the case of
V471\,Tau the optical emission is largely dominated by the K2V
companion, and with $B-V\simeq0.76$ the system does clearly not
qualify as a white dwarf candidate.

\subsection{\label{s-lackofearlysecs}The lack of early companions}
So far we have focussed much of our attention on the temperatures of
the white dwarfs in our PCEB sample, as this parameter is a direct
measure of the age of these systems. However, Table\,\ref{t-sample}
suggests that this sample is also biased with respect to the
\textit{mass of the companion stars}. The average mass of the
companion in our PCEB sample is
$\overline{M}_{\mathrm{sec~PCEB}}=0.24$, whereas CVcat
\citep{kubeetal02-1} lists the masses for 80 CVs with
$\overline{M}_{\mathrm{sec~CV}}=0.43$. A direct consequence of this
dominance of low mass companion stars in the known PCEB sample is that
the majority of these systems evolve into contact at short orbital
periods (Sect.\,\ref{s-future}; Fig.\,\ref{f-cvporb}). 

The shortage of PCEBs with massive/early-type donor stars
($0.4\,\Msun\la\Msec\la0.8\,\Msun$), i.e. the progenitors of
long-period CVs, is subject to the selection effect which we just
discussed in Sect.\,\ref{s-finding_pcebs}: in PCEBs with an early-type
companion the secondary contributes significantly to the total optical
emission of the system, and, consequently, such binaries do not
qualify as white dwarf candidates (Fig.\,\ref{f-pceb_colours}), the
major discovery channel for PCEBs.

As a result of the (very likely) incompleteness of known PCEBs with an
early-type donor the space density estimates presented in
Sect.\,\ref{s-space_densities} should be considered to be lower limits
to the true $\rho_{\mathrm{PCEB}}$ and $\rho_{\mathrm{preCV}}$. 

\section{Conclusion}
We have presented the properties of the currently known sample of
well-observed PCEBs and have calculated their past and future
evolution based on two different assumptions for the angular
momentum loss mechanism, adopting either the ``classical'' disrupted
magnetic braking prescription (\Jcmb) or the AML prescription derived
by \citet{sillsetal00-1} (\Jrmb). The results from this study are:

\begin{enumerate}
\item The presently known PCEB population is dominated by young
systems most of which have evolved only through a small fraction of
their lifetime as detached binaries. On average, the white dwarfs in
PCEBs are hotter than field white dwarfs.

\item While the evolution of an individual system strongly depends on
whether we assume \Jrmb or \Jcmb, 
the total number of pre-CVs within the PCEB sample, i.e. systems which will 
start mass transfer within a Hubble time, is nearly the same for both 
prescriptions. Considering the time scale on which the PCEBs will evolve into 
semi-detached CVs we predict the existence
of a large population of old PCEBs containing cold white dwarfs. The
present lack of such systems is very likely the result of
observational biases, as the majority of the known PCEBs were
initially selected as blue objects.

\item An additional consequence of the observational selection effects
involved in the discovery of PCEBs is a shortage of systems containing 
a ``massive''/early-type companion
($0.4\,\Msun\la\Msec\la0.8\,\Msun$)~--~which are the progenitors of
long-period CVs. Indeed, our calculations predict that most of the
pre-CVs among the PCEBs in our sample will evolve into CV with short
orbital periods ($\Porb<3$\,h). 

\item The space density of PCEBs estimated from the currently known
sample is $\,6\times10^{-6}\mathrm{pc}^{-3} \lappr \rho_{\mathrm{PCEB}}
\lappr 3\times\,10^{-5}\mathrm{pc}^{-3}$, depending on the assumed AML
prescription.  Taking into account the age as well as the evolution
time scale of the pre-CVs in our sample and assuming that the
birthrate of PCEBs remains unchanged we derive for the CV space
density $\rho_{\mathrm{CV}}\sim\,10^{-5}\mathrm{pc}^{-3}$, nearly independent
on whether we assume \Jcmb or \Jrmb.  This value is somewhat higher
than current observational estimates but below theoretical
predictions.  However, due to the observational bias working against
the discovery of PCEBs containing an early-type companion our
estimates should be considered to be lower limits to the true values
of $\rho_{\mathrm{CV}}$ and $\rho_{\mathrm{PCEB}}$.

\end{enumerate}

\begin{acknowledgement}
We thank for support by an individual Marie-Curie Fellowship (MRS) and
a PPARC Advanced Fellowship (BTG).
\end{acknowledgement}

\noindent
{\bf Note added in proof:} After the submission of our paper,
\citet{raymondetal03-1} have published a list of 109 PCEBs found in
the Sloan Digital Sky Survey. The average white dwarf temperature of
this sample 16\,000\,K, significantly lower than that of our sample of
PCEBs with known orbital period (Table\,\ref{t-sample}), and
demonstrates that more sophisticated selection criteria (compared to
pure ``blue'' surveys) can successfully identify PCEBs containing rather
cold white dwarfs. Unfortunately, only a single system in the
list of \citet{raymondetal03-1} has an orbital period
measurement~--~determining the periods of their PCEBs harbouring white
dwarfs with $\Twd\la12000$\,K provides a direct test for our
prediction (2) above.

\appendix

\section{The discovery history of our PCEB sample (Table\,1)}

\textit{RR\,Cae:} Discovered as high proper motion object by
\citet{luyten55-1}, DA white dwarf plus Balmer emission line spectrum
noted by \citet{rodgers+eggen74-1}, late-type nature for the companion
suggested by \citet{bessell+wickramasinghe79-1}, discovery of eclipses
and orbital period measurement (photometric) by
\citet{krzeminski84-1}. Two comprehensive studies by
\citet{bruch+diaz98-1} and \citet{bruch99-1}.

\textit{EG\,UMa:} Discovered as a white dwarf on Schmidt prism plates
by \citet{stephenson60-1}, emission lines noted by
\citet{greenstein65-1}, orbital period (spectroscopic) by
\citet{lanning82-1}. Recent comprehensive study by
\citet{bleachetal00-1}.

\textit{EC\,13471--1258:} Discovered in the Edinburgh-Cape faint blue
object survey of high galactic latitudes, binarity revealed through
the detection of eclipses \citep{kilkennyetal97-1}.
Recent study by \citet{kawkaetal02-1}.

\textit{BPM\,71214:} Discovered as a high proper motion object
\citep{luyten63-1}. 
Recent study by \citet{kawkaetal02-1}.

\textit{HR\,Cam:} Listed as WD candidate by \citet{giclasetal70-1},
spectroscopic confirmation of WD nature by \citet{wills+wills74-1},
detection of a red companion by \citet{zuckerman+becklin92-1}, orbital
period (spectroscopic and photometric) from the comprehensive studies
of \citet{marsh+duck96-1,maxtedetal98-1}.

\textit{UZ\,Sex:} Discovered in the PG survey, listed as a
DA/composite spectral type by \citet{greenetal86-1}, orbital period
(spectroscopic) from the comprehensive study of
\citet{safferetal93-1}, see also the recent analyses by
\cite{bruch+diaz99-1, bleachetal00-1}.

\textit{BPM\,6502:} Discovered as a high proper motion star by
\citet{luyten57-1}, listed as WD candidate by \citet{eggen69-1},
spectroscopically confirmed as a WD by \citet{wegner73-1}, detection
of companion and orbital period (spectroscopic) by
\citet{kawkaetal00-1}. 

\textit{HZ\,9:} Discovered in a search for faint blue stars by
\citet{humason+zwicky47-1}, mentioned as a DA+dMe candidate by
\citet{greenstein58-1}, orbital period (spectroscopic) measured by
\citet{lanning+pesch81-1}. 

\textit{MS\,Peg:} Listed as WD candidate by \citet{giclasetal65-1},
spectroscopic confirmation of WD nature by \citet{greenstein69-1},
emission lines and radial velocity variations detected by
\citet{tytler+rubenstein89-1} and \citet{schultzetal93-1}, orbital
period (spectroscopic) measured by \cite{schmidtetal95-3}.

\textit{HS\,0705+6700:} Classified in the Hamburg Quasar Survey
\citep{hagenetal95-1} as hot star candidate, eclipses discovered and
orbital period measured by \citet{drechseletal01-1}.

\textit{HW\,Vir:} Listed as ultraviolet-bright star by
\citet{carnochan+wilson83-1} and classified as likely sdB star by
\citet{berger+fringant80-1}, discovery of eclipses and orbital period
(photometric) by \citet{menzies86-1}.

\textit{LM\,Com:} Listed as blue object by
\citet{iriarte+chavira57-1}, also detected in the PG survey
\citep{greenetal86-1}, and identified as DA+dM binary by
\citet{fergusonetal84-1}, \Ha\ emission detected by
\citet{oroszetal97-1}, orbital period (spectroscopic) and
comprehensive study by \citet{oroszetal99-1}.

\textit{PG\,1017--086:} Discovered in the PG survey, classified as 
sdB star by \citet{greenetal86-1}, variability  discovered and orbital
period measured by \citet{maxtedetal02-1}.

\textit{CC\,Cet:}  Discovered in the PG survey, listed as a
DA/composite spectral type by \citet{greenetal86-1}, orbital period
(spectroscopic) from the comprehensive study of
\citet{safferetal93-1}, see also \citet{somersetal96-2}. 

\textit{V471\,Tau:} The brightest and  best-studied pre-CV. Listed as a
spectroscopic  binary by  \citet{wilson53-1},  discovery of  eclipses,
spectral  classification  as  DA+dK  binary, and  measurement  of  the
orbital period (photometric) by \citet{nelson+young70-1}.

\textit{PG\,1336--018:} Discovered in the PG survey, classified as 
sdB star by \citet{greenetal86-1}, eclipses discovered and orbital
period measured by \citet{kilkennyetal98-1}.

\textit{AA\,Dor:} Listed as faint blue star by \citet{luyten57-2} and
as spectroscopically variable object in the foreground of the LMC
by \citet{feastetal60-1}, discovery of eclipses and measurement of the
orbital period (photometric) by \citet{kilkennyetal78-1}.

\textit{RE\,2013+400:} Discovered as bright EUV source during the
ROSAT Wide Field Camera all-sky survey and listed as a WD by
\citet{poundsetal93-1}, variable Balmer emission was noted by
\citet{barstowetal93-1}, the orbital period (spectroscopic) was
measured by \citet{thorstensenetal94-1}.

\textit{GK\,Vir:} Discovered in the PG survey, eclipse discovered and
orbital period (photometric) measured by \citet{greenetal78-1}.

\textit{MT\,Ser: } Classified as PN on the Palomar Sky Survey plates
by \citet{abell55-1}, variability suggested by \citet{abell66-1},
recurrent photometric variability was detected by
\citet{grauer+bond83-1} who also suggested the sdO+dM nature of the
system, for a more comprehensive photometric study, see
\citet{bruchetal01-1}.

\textit{IM\,CMa:} Discovered as bright EUV source during the
ROSAT Wide Field Camera all-sky survey and listed as a WD by
\citet{poundsetal93-1}, Balmer emission detected and orbital period
(spectroscopic) measured by \citet{vennes+thorstensen94-1}. 

\textit{NN\,Ser:} Discovered as a CV candidate in the PG survey by
\citet{greenetal86-1}, discovery of deep eclipses, classification as
DA+dM binary and measurement of the orbital period (photometric) by
\citet{haefner89-1}. 

\textit{RE\,1016--053:} Discovered as bright EUV source during the
ROSAT Wide Field Camera all-sky survey and listed as a WD by
\citet{poundsetal93-1}, classified as DA+dM by
\citet{jomaronetal93-1},  orbital period (spectroscopic) 
measured by \citet{tweedyetal93-1}.

\textit{V477\,Lyr:} Classified as PN on the Palomar Sky Survey plates by
\citet{abell55-1},  variability suggested by \citet{abell66-1},
eclipse discovered and period (photometric) measured by \cite{bond80-1}.

\textit{Abell\,65:} Classified as PN on the Palomar Sky Survey plates
by \citet{abell66-1}, an estimate of the orbital period (photometric)
was published by \cite{bond+livio90-1}.

\textit{KV\,Vel:} Listed as luminous star in the list of
\citet{stephenson+sanduleak71-1}, planetary nebula nature suggested by
\citet{holmbergetal78-1}, spectroscopically identified as sdO star
within a planetary nebula by \citet{drilling83-1}. Photometric
variability discovered and orbital period (photometric) measured by
\cite{drilling+bravo84-1}. 

\textit{UU\,Sge:} Classified as PN on the Palomar Sky Survey plates by
\citet{abell55-1}, variability suggested by \citet{abell66-1},
eclipses discovered and orbital period (photometric) measured by
\citet{milleretal76-1}.

\textit{HS\,1136+6646:}  Classified in the Hamburg Quasar Survey
\citep{hagenetal95-1} as hot star candidate, identified as a
spectroscopic binary by \citet{heberetal96-1}, orbital period
(spectroscopic) measured by \citet{singetal01-1}.

\textit{BE\,UMa:} Identified as variable star and orbital period
(photometric) by \citet{kurochkin64-1,kurochkin71-1}, recovered as
emission line star in the PG survey by \citet{fergusonetal81-1}.


\begin{thebibliography}{161}
\expandafter\ifx\csname natexlab\endcsname\relax\def\natexlab#1{#1}\fi

\bibitem[{{Abell}(1955)}]{abell55-1}
{Abell}, G.~O. 1955, PASP, 67, 258

\bibitem[{{Abell}(1966)}]{abell66-1}
---. 1966, ApJ, 144, 259

\bibitem[{{Althaus} \& {Benvenuto}(1997)}]{althaus+benvenuto97-1}
{Althaus}, L.~G. \& {Benvenuto}, O.~G. 1997, ApJ, 477, 313

\bibitem[{{Andronov} {et~al.}(2003){Andronov}, {Pinsonneault}, \&
  {Sills}}]{andronovetal03-1}
{Andronov}, N., {Pinsonneault}, M., \& {Sills}, A. 2003, ApJ, 582, 358

\bibitem[{{Barstow} {et~al.}(1993){Barstow}, {Hodgkin}, {Pye}, {King},
  {Fleming}, {Holberg}, \& {Tweedy}}]{barstowetal93-1}
{Barstow}, M.~A., {Hodgkin}, S.~T., {Pye}, J.~P., {et~al.} 1993, in White
  Dwarfs: Advances in Observation and Theory, ed. M.~A. {Barstow}, NATO ASIC
  Proc. No. 403 (Dordrecht: Kluwer), 433

\bibitem[{{Bell} {et~al.}(1994){Bell}, {Pollacco}, \&
  {Hilditch}}]{belletal94-1}
{Bell}, S.~A., {Pollacco}, D.~L., \& {Hilditch}, R.~W. 1994, MNRAS, 270, 449

\bibitem[{{Berger} \& {Fringant}(1980)}]{berger+fringant80-1}
{Berger}, J. \& {Fringant}, A.-M. 1980, A\&A, 85, 367

\bibitem[{{Bergeron} {et~al.}(1994){Bergeron}, {Wesemael}, {Beauchamp}, {Wood},
  {Lamontagne}, {Fontaine}, \& {Liebert}}]{bergeronetal94-1}
{Bergeron}, P., {Wesemael}, F., {Beauchamp}, A., {et~al.} 1994, ApJ, 432, 305

\bibitem[{{Bessell} \& {Wickramasinghe}(1979)}]{bessell+wickramasinghe79-1}
{Bessell}, M.~S. \& {Wickramasinghe}, D.~T. 1979, ApJ, 227, 232

\bibitem[{{Bleach} {et~al.}(2000){Bleach}, {Wood}, {Catal{\' a}n}, {Welsh},
  {Robinson}, \& {Skidmore}}]{bleachetal00-1}
{Bleach}, J.~N., {Wood}, J.~H., {Catal{\' a}n}, M.~S., {et~al.} 2000, MNRAS,
  312, 70

\bibitem[{{Bond}(1980)}]{bond80-1}
{Bond}, H.~E. 1980, IAU Circ., 3480

\bibitem[{{Bond} \& {Livio}(1990)}]{bond+livio90-1}
{Bond}, H.~E. \& {Livio}, M. 1990, ApJ, 355, 568

\bibitem[{{Bragaglia} {et~al.}(1995){Bragaglia}, {Renzini}, \&
  {Bergeron}}]{bragagliaetal95-1}
{Bragaglia}, A., {Renzini}, A., \& {Bergeron}, P. 1995, ApJ, 443, 735

\bibitem[{{Bruch}(1999)}]{bruch99-1}
{Bruch}, A. 1999, AJ, 117, 3031

\bibitem[{{Bruch} \& {Diaz}(1998)}]{bruch+diaz98-1}
{Bruch}, A. \& {Diaz}, M.~P. 1998, AJ, 116, 908

\bibitem[{{Bruch} \& {Diaz}(1999)}]{bruch+diaz99-1}
---. 1999, A\&A, 351, 573

\bibitem[{{Bruch} {et~al.}(2001){Bruch}, {Vaz}, \& {Diaz}}]{bruchetal01-1}
{Bruch}, A., {Vaz}, L.~P.~R., \& {Diaz}, M.~P. 2001, A\&A, 377, 898

\bibitem[{{Cahn} \& {Kaler}(1971)}]{cahn+kaler71-1}
{Cahn}, J.~H. \& {Kaler}, J.~B. 1971, ApJS, 22, 319

\bibitem[{{Cahn} {et~al.}(1992){Cahn}, {Kaler}, \&
  {Stanghellini}}]{cahnetal92-1}
{Cahn}, J.~H., {Kaler}, J.~B., \& {Stanghellini}, L. 1992, A\&AS, 94, 399

\bibitem[{{Carnochan} \& {Wilson}(1983)}]{carnochan+wilson83-1}
{Carnochan}, D.~J. \& {Wilson}, R. 1983, MNRAS, 202, 317

\bibitem[{{Catalan} {et~al.}(1994){Catalan}, {Davey}, {Sarna}, {Connon-Smith},
  \& {Wood}}]{catalanetal94-1}
{Catalan}, M.~S., {Davey}, S.~C., {Sarna}, M.~J., {Connon-Smith}, R., \&
  {Wood}, J.~H. 1994, MNRAS, 269, 879

\bibitem[{{Cayrel} {et~al.}(2001){Cayrel}, {Hill}, {Beers}, {Barbuy}, {Spite},
  {Spite}, {Plez}, {Andersen}, {Bonifacio}, {Fran{\c c}ois}, {Molaro},
  {Nordstr{\" o}m}, \& {Primas}}]{cayreletal01-1}
{Cayrel}, R., {Hill}, V., {Beers}, T.~C., {et~al.} 2001, Nat, 409, 691

\bibitem[{{Chen} {et~al.}(1995){Chen}, {O'Donoghue}, {Stobie}, {Kilkenny},
  {Roberts}, \& {Van Wyk}}]{chenetal95-1}
{Chen}, A., {O'Donoghue}, D., {Stobie}, R.~S., {et~al.} 1995, MNRAS, 275, 100

\bibitem[{{Clemens} {et~al.}(1998){Clemens}, {Reid}, {Gizis}, \&
  {O'Brien}}]{clemensetal98-1}
{Clemens}, J.~C., {Reid}, I.~N., {Gizis}, J.~E., \& {O'Brien}, M.~S. 1998, ApJ,
  496, 352

\bibitem[{{de Kool}(1992)}]{dekool92-1}
{de Kool}, M. 1992, A\&A, 261, 188

\bibitem[{{de Kool} \& {Ritter}(1993)}]{dekool+ritter93-1}
{de Kool}, M. \& {Ritter}, H. 1993, A\&A, 267, 397

\bibitem[{{Downes}(1986)}]{downes86-1}
{Downes}, R.~A. 1986, ApJ, 307, 170

\bibitem[{{Drechsel} {et~al.}(2001){Drechsel}, {Heber}, {Napiwotzki},
  {{\O}stensen}, {Solheim}, {Johannessen}, {Schuh}, {Deetjen}, \&
  {Zola}}]{drechseletal01-1}
{Drechsel}, H., {Heber}, U., {Napiwotzki}, R., {et~al.} 2001, A\&A, 379, 893

\bibitem[{{Driebe} {et~al.}(1999){Driebe}, {Bl\"ocker}, {Sch\"onberner}, \&
  {Herwig}}]{driebeetal99-1}
{Driebe}, T., {Bl\"ocker}, T., {Sch\"onberner}, D., \& {Herwig}, F. 1999, A\&A,
  350, 89

\bibitem[{{Driebe} {et~al.}(1998){Driebe}, {Schoenberner}, {Bloecker}, \&
  {Herwig}}]{driebeetal98-1}
{Driebe}, T., {Schoenberner}, D., {Bloecker}, T., \& {Herwig}, F. 1998, A\&A,
  339, 123

\bibitem[{{Drilling}(1983)}]{drilling83-1}
{Drilling}, J.~S. 1983, ApJ Lett., 270, L13

\bibitem[{{Drilling} \& {Bravo}(1984)}]{drilling+bravo84-1}
{Drilling}, J.~S. \& {Bravo}, J. 1984, IAU Circ., 3939

\bibitem[{{Dubus} {et~al.}(2002){Dubus}, {Taam}, \& {Spruit}}]{dubusetal02-1}
{Dubus}, G., {Taam}, R.~E., \& {Spruit}, H.~C. 2002, ApJ, 569, 395

\bibitem[{{Eggen}(1969)}]{eggen69-1}
{Eggen}, O.~J. 1969, ApJ, 157, 287

\bibitem[{{Eggleton}(1983)}]{eggleton83-1}
{Eggleton}, P.~P. 1983, ApJ, 268, 368

\bibitem[{{Feast} {et~al.}(1960){Feast}, {Thackeray}, \&
  {Wesselink}}]{feastetal60-1}
{Feast}, M.~W., {Thackeray}, A.~D., \& {Wesselink}, A.~J. 1960, MNRAS, 121, 337

\bibitem[{{Ferguson} {et~al.}(1984){Ferguson}, {Green}, \&
  {Liebert}}]{fergusonetal84-1}
{Ferguson}, D.~H., {Green}, R.~F., \& {Liebert}, J. 1984, ApJ, 287, 320

\bibitem[{{Ferguson} {et~al.}(1999){Ferguson}, {Liebert}, {Haas}, {Napiwotzki},
  \& {James}}]{fergusonetal99-1}
{Ferguson}, D.~H., {Liebert}, J., {Haas}, S., {Napiwotzki}, R., \& {James},
  T.~A. 1999, ApJ, 518, 866

\bibitem[{{Ferguson} {et~al.}(1981){Ferguson}, {McGraw}, {Spinrad}, {Liebert},
  \& {Green}}]{fergusonetal81-1}
{Ferguson}, D.~H., {McGraw}, J.~T., {Spinrad}, H., {Liebert}, J., \& {Green},
  R.~F. 1981, ApJ, 251, 205

\bibitem[{{Ferreras} {et~al.}(2001){Ferreras}, {Melchiorri}, \&
  {Silk}}]{ferrerasetal01-1}
{Ferreras}, I., {Melchiorri}, A., \& {Silk}, J. 2001, MNRAS, 327, L47

\bibitem[{{Fontaine} {et~al.}(2001){Fontaine}, {Brassard}, \&
  {Bergeron}}]{fontaineetal01-1}
{Fontaine}, G., {Brassard}, P., \& {Bergeron}, P. 2001, PASP, 113, 409

\bibitem[{{Fujimoto} \& {Iben}(1989)}]{fujimoto+iben89-1}
{Fujimoto}, M.~Y. \& {Iben}, I.~J. 1989, ApJ, 341, 306

\bibitem[{{Fulbright} {et~al.}(1993){Fulbright}, {Liebert}, {Bergeron}, \&
  {Green}}]{fulbrightetal93-1}
{Fulbright}, M.~S., {Liebert}, J., {Bergeron}, P., \& {Green}, R. 1993, ApJ,
  406, 240

\bibitem[{{G\"ansicke}(2000)}]{gaensicke00-1}
{G\"ansicke}, B.~T. 2000, Reviews of Modern Astronomy, 13, 151

\bibitem[{{G\"ansicke} {et~al.}(1995){G\"ansicke}, {Beuermann}, \& {de
  Martino}}]{gaensickeetal95-1}
{G\"ansicke}, B.~T., {Beuermann}, K., \& {de Martino}, D. 1995, A\&A, 303, 127

\bibitem[{{G\"ansicke} {et~al.}(2002){G\"ansicke}, {Hagen}, \&
  {Engels}}]{gaensickeetal02-2}
{G\"ansicke}, B.~T., {Hagen}, H.~J., \& {Engels}, D. 2002, On the
  space density of cataclysmic variables, ed. B.~T. {G\"ansicke},
  K.~{Beuermann}, \& K.~{Reinsch} (ASP Conf. Ser. 261), 190--199

\bibitem[{{Giclas} {et~al.}(1965){Giclas}, {Burnham}, \&
  {Thomas}}]{giclasetal65-1}
{Giclas}, H.~L., {Burnham}, R., \& {Thomas}, N.~G. 1965, Lowell Obs. Bull.,
  125, 155

\bibitem[{{Giclas} {et~al.}(1970){Giclas}, {Burnham}, \&
  {Thomas}}]{giclasetal70-1}
---. 1970, Lowell Obs. Bull., 153, 183

\bibitem[{{Grauer} \& {Bond}(1983)}]{grauer+bond83-1}
{Grauer}, A.~D. \& {Bond}, H.~E. 1983, ApJ, 271, 259

\bibitem[{{Green} {et~al.}(1984){Green}, {Liebert}, \&
  {Wesemael}}]{greenetal84-1}
{Green}, R.~F., {Liebert}, J., \& {Wesemael}, F. 1984, ApJ, 280, 177

\bibitem[{{Green} {et~al.}(1978){Green}, {Richstone}, \&
  {Schmidt}}]{greenetal78-1}
{Green}, R.~F., {Richstone}, D.~O., \& {Schmidt}, M. 1978, ApJ, 224, 892

\bibitem[{{Green} {et~al.}(1986){Green}, {Schmidt}, \&
  {Liebert}}]{greenetal86-1}
{Green}, R.~F., {Schmidt}, M., \& {Liebert}, J. 1986, ApJS, 61, 305

\bibitem[{{Greenstein}(1958)}]{greenstein58-1}
{Greenstein}, J.~L. 1958, Handbuch der Physik, Vol.~50 (Berlin: Springer), 161

\bibitem[{{Greenstein}(1965)}]{greenstein65-1}
{Greenstein}, J.~L. 1965, in First Conference on Faint Blue Stars, ed.
  W.~{Luyten} (Minneapolis: University of Minnesota Press), 97

\bibitem[{{Greenstein}(1969)}]{greenstein69-1}
---. 1969, ApJ, 158, 281

\bibitem[{{Guinan} \& {Sion}(1984)}]{guinan+sion84-1}
{Guinan}, E.~F. \& {Sion}, E.~M. 1984, AJ, 89, 1252

\bibitem[{{Haefner}(1989)}]{haefner89-1}
{Haefner}, R. 1989, A\&A, 213, L15

\bibitem[{{Hagen} {et~al.}(1995){Hagen}, {Groote}, {Engels}, \&
  {Reimers}}]{hagenetal95-1}
{Hagen}, H.~J., {Groote}, D., {Engels}, D., \& {Reimers}, D. 1995, A\&AS, 111,
  195

\bibitem[{{Heber} {et~al.}(1996){Heber}, {Dreizler}, \&
  {Hagen}}]{heberetal96-1}
{Heber}, U., {Dreizler}, S., \& {Hagen}, H.-J. 1996, A\&A, 311, L17

\bibitem[{{Herrero} {et~al.}(1990){Herrero}, {Manchado}, \&
  {Mendez}}]{herreroetal90-1}
{Herrero}, A., {Manchado}, A., \& {Mendez}, R.~H. 1990, Ap\&SS, 169, 183

\bibitem[{{Hilditch} {et~al.}(1996){Hilditch}, {Harries}, \&
  {Hill}}]{hilditchetal96-1}
{Hilditch}, R.~W., {Harries}, T.~J., \& {Hill}, G. 1996, MNRAS, 279, 1380

\bibitem[{{Holmberg} {et~al.}(1978){Holmberg}, {Lauberts}, {Schuster}, \&
  {West}}]{holmbergetal78-1}
{Holmberg}, E.~B., {Lauberts}, A., {Schuster}, H.~E., \& {West}, R.~M. 1978,
  A\&AS, 31, 15

\bibitem[{{Howell} {et~al.}(2001){Howell}, {Nelson}, \&
  {Rappaport}}]{howelletal01-1}
{Howell}, S.~B., {Nelson}, L.~A., \& {Rappaport}, S. 2001, ApJ, 550, 897

\bibitem[{{Howell} {et~al.}(1997){Howell}, {Rappaport}, \&
  {Politano}}]{howelletal97-1}
{Howell}, S.~B., {Rappaport}, S., \& {Politano}, M. 1997, MNRAS, 287, 929

\bibitem[{{Humason} \& {Zwicky}(1947)}]{humason+zwicky47-1}
{Humason}, M.~L. \& {Zwicky}, F. 1947, ApJ, 105, 85

\bibitem[{{Iben} \& {Tutukov}(1993)}]{iben+tutukov93-1}
{Iben}, I., J. \& {Tutukov}, A.~V. 1993, ApJ, 418, 343

\bibitem[{{Iriarte} \& {Chavira}(1957)}]{iriarte+chavira57-1}
{Iriarte}, B. \& {Chavira}, E. 1957, Boletin de los Observatorios Tonantzintla
  y Tacubaya, 2, 3

\bibitem[{{Jomaron} {et~al.}(1993){Jomaron}, {Raymont}, {Bromage}, {Hassall},
  {Hodgkin}, {Mason}, {Naylor}, \& {Watson}}]{jomaronetal93-1}
{Jomaron}, C.~M., {Raymont}, G.~B., {Bromage}, G.~E., {et~al.} 1993, MNRAS,
  264, 219

\bibitem[{{Kawka} {et~al.}(2000){Kawka}, {Vennes}, {Dupuis}, \&
  {Koch}}]{kawkaetal00-1}
{Kawka}, A., {Vennes}, S.~., {Dupuis}, J., \& {Koch}, R. 2000, AJ, 120, 3250

\bibitem[{{Kawka} {et~al.}(2002){Kawka}, {Vennes}, {Koch}, \&
  {Williams}}]{kawkaetal02-1}
{Kawka}, A., {Vennes}, S., {Koch}, R., \& {Williams}, A. 2002, AJ, 124, 2853

\bibitem[{{Kepler} \& {Nelan}(1993)}]{kepler+nelan93-1}
{Kepler}, S.~O. \& {Nelan}, E.~P. 1993, AJ, 105, 608

\bibitem[{{Kilkenny} {et~al.}(1978){Kilkenny}, {Hilditch}, \&
  {Penfold}}]{kilkennyetal78-1}
{Kilkenny}, D., {Hilditch}, R.~W., \& {Penfold}, J.~E. 1978, MNRAS, 183, 523

\bibitem[{{Kilkenny} {et~al.}(1998){Kilkenny}, {O'Donoghue}, {Koen},
  {Lynas-Gray}, \& {van Wyk}}]{kilkennyetal98-1}
{Kilkenny}, D., {O'Donoghue}, D., {Koen}, C., {Lynas-Gray}, A.~E., \& {van
  Wyk}, F. 1998, MNRAS, 296, 329

\bibitem[{{Kilkenny} {et~al.}(1997){Kilkenny}, {O'Donoghue}, {Koen}, {Stobie},
  \& {Chen}}]{kilkennyetal97-1}
{Kilkenny}, D., {O'Donoghue}, D., {Koen}, C., {Stobie}, R.~S., \& {Chen}, A.
  1997, MNRAS, 287, 867

\bibitem[{{King}(1988)}]{king88-1}
{King}, A.~R. 1988, QJRAS, 29, 1

\bibitem[{{King} {et~al.}(1995){King}, {Frank}, {Kolb}, \&
  {Ritter}}]{kingetal95-1}
{King}, A.~R., {Frank}, J., {Kolb}, U., \& {Ritter}, H. 1995, ApJ Lett., 444,
  L37

\bibitem[{{King} \& {Kolb}(1995)}]{king+kolb95-1}
{King}, A.~R. \& {Kolb}, U. 1995, ApJ, 439, 330

\bibitem[{{King} {et~al.}(1994){King}, {Kolb}, {De Kool}, \&
  {Ritter}}]{kingetal94-1}
{King}, A.~R., {Kolb}, U., {De Kool}, M., \& {Ritter}, H. 1994, MNRAS, 269, 907

\bibitem[{{King} {et~al.}(2002){King}, {Osborne}, \& {Schenker}}]{kingetal02-2}
{King}, A.~R., {Osborne}, J.~P., \& {Schenker}, K. 2002, MNRAS, 329, L43

\bibitem[{{King} \& {Schenker}(2002)}]{king+schenker02-1}
{King}, A.~R. \& {Schenker}, K. 2002, in The Physics of Cataclysmic Variables
  and Related Objects, ed. B.~T. {G\"ansicke}, K.~{Beuermann}, \& K.~{Reinsch}
  (ASP Conf. Ser. 261), 233--241

\bibitem[{{Koester}(2002)}]{koester02-1}
{Koester}, D. 2002, A\&AR, 11, 33

\bibitem[{{Koester} {et~al.}(1979){Koester}, {Schulz}, \&
  {Weidemann}}]{koesteretal79-1}
{Koester}, D., {Schulz}, H., \& {Weidemann}, V. 1979, A\&A, 76, 262

\bibitem[{{Kolb}(2002)}]{kolb02-1}
{Kolb}, U. 2002, in The Physics of Cataclysmic Variables and Related Objects,
  ed. B.~T. {G\"ansicke}, K.~{Beuermann}, \& K.~{Reinsch} (ASP Conf. Ser. 261),
  180--189

\bibitem[{{Kolb} \& {Baraffe}(1999)}]{kolb+baraffe99-1}
{Kolb}, U. \& {Baraffe}, I. 1999, MNRAS, 309, 1034

\bibitem[{{Kolb} {et~al.}(1998){Kolb}, {King}, \& {Ritter}}]{kolbetal98-1}
{Kolb}, U., {King}, A.~R., \& {Ritter}, H. 1998, MNRAS, 298, L29

\bibitem[{{Krishnamurthi} {et~al.}(1997){Krishnamurthi}, {Pinsonneault},
  {Barnes}, \& {Sofia}}]{krishnamurthietal97-1}
{Krishnamurthi}, A., {Pinsonneault}, M.~H., {Barnes}, S., \& {Sofia}, S. 1997,
  ApJ, 480, 303

\bibitem[{{Krzeminski}(1984)}]{krzeminski84-1}
{Krzeminski}, W. 1984, IAU Circ., 4014

\bibitem[{{Kube} {et~al.}(2002){Kube}, {G\"ansicke}, \&
  {Hoffmann}}]{kubeetal02-1}
{Kube}, J., {G\"ansicke}, B., \& {Hoffmann}, B. 2002, in The Physics of
  Cataclysmic Variables and Related Objects, ed. B.~T. {G\"ansicke},
  K.~{Beuermann}, \& K.~{Reinsch} (ASP Conf. Ser. 261), 678--679

\bibitem[{{Kudritzki} {et~al.}(1982){Kudritzki}, {Simon}, {Lynas-Gray},
  {Kilkenny}, \& {Hill}}]{kudritzkietal82-1}
{Kudritzki}, R.~P., {Simon}, K.~P., {Lynas-Gray}, A.~E., {Kilkenny}, D., \&
  {Hill}, P.~W. 1982, A\&A, 106, 254

\bibitem[{{Kurochkin}(1964)}]{kurochkin64-1}
{Kurochkin}, N.~E. 1964, Peremnye Zvezdy (Academy Sci. U.S.S.R., Var. Star
  Bull), 15, 77

\bibitem[{{Kurochkin}(1971)}]{kurochkin71-1}
---. 1971, Peremnye Zvezdy (Academy Sci. U.S.S.R., Var. Star Bull), 18, 85

\bibitem[{{Lanning}(1982)}]{lanning82-1}
{Lanning}, H.~H. 1982, ApJ, 253, 752

\bibitem[{{Lanning} \& {Pesch}(1981)}]{lanning+pesch81-1}
{Lanning}, H.~H. \& {Pesch}, P. 1981, ApJ, 244, 280

\bibitem[{{Livio} \& {Pringle}(1994)}]{livio+pringle94-1}
{Livio}, M. \& {Pringle}, J.~E. 1994, ApJ, 427, 956

\bibitem[{{Luyten}(1955)}]{luyten55-1}
{Luyten}, W.~J. 1955, {Luyten's Five Tenths.} (Minneapolis: Lund Press)

\bibitem[{{Luyten}(1957{\natexlab{a}})}]{luyten57-1}
---. 1957{\natexlab{a}}, {A catalogue of 9867 stars in the Southern Hemisphere
  with proper motions exceeding 0."2 annually.} (Minneapolis: Lund Press)

\bibitem[{{Luyten}(1957{\natexlab{b}})}]{luyten57-2}
---. 1957{\natexlab{b}}, A Search for Faint Blue Stars (Minneapolis: University
  of Minnesota)

\bibitem[{{Luyten}(1963)}]{luyten63-1}
---. 1963, Bruce proper motion survey. The general catalogue. Vol. I, II.

\bibitem[{{Maciel}(1984)}]{maciel84-1}
{Maciel}, W.~J. 1984, A\&AS, 55, 253

\bibitem[{{Marsh} {et~al.}(1995){Marsh}, {Dhillon}, \& {Duck}}]{marshetal95-1}
{Marsh}, T.~R., {Dhillon}, V.~S., \& {Duck}, S.~R. 1995, MNRAS, 275, 828

\bibitem[{{Marsh} \& {Duck}(1996)}]{marsh+duck96-1}
{Marsh}, T.~R. \& {Duck}, S.~R. 1996, MNRAS, 278, 565

\bibitem[{{Maxted} {et~al.}(2002){Maxted}, {Marsh}, {Heber}, {Morales-Rueda},
  {North}, \& {Lawson}}]{maxtedetal02-1}
{Maxted}, P.~F.~L., {Marsh}, T.~R., {Heber}, U., {et~al.} 2002, MNRAS, 333, 231

\bibitem[{{Maxted} {et~al.}(1998){Maxted}, {Marsh}, {Moran}, {Dhillon}, \&
  {Hilditch}}]{maxtedetal98-1}
{Maxted}, P. F.~L., {Marsh}, T.~R., {Moran}, C., {Dhillon}, V.~S., \&
  {Hilditch}, R.~W. 1998, MNRAS, 300, 1225

\bibitem[{{McCook} \& {Sion}(1999)}]{mccook+sion99-1}
{McCook}, G.~P. \& {Sion}, E.~M. 1999, ApJS, 121, 1

\bibitem[{{Menzies}(1986)}]{menzies86-1}
{Menzies}, J.~W. 1986, Ann. Rep. SAAO 1985, 20

\bibitem[{{Miller} {et~al.}(1976){Miller}, {Krzeminski}, \&
  {Priedhorsky}}]{milleretal76-1}
{Miller}, J.~S., {Krzeminski}, W., \& {Priedhorsky}, W. 1976, IAU Circ., 2974

\bibitem[{{Nelson} \& {Young}(1970)}]{nelson+young70-1}
{Nelson}, B. \& {Young}, A. 1970, PASP, 82, 699

\bibitem[{{O'Brien} {et~al.}(2001){O'Brien}, {Bond}, \&
  {Sion}}]{obrienetal01-1}
{O'Brien}, M.~S., {Bond}, H.~E., \& {Sion}, E.~M. 2001, ApJ, 563, 971

\bibitem[{{Orosz} {et~al.}(1997){Orosz}, {Wade}, \& {Harlow}}]{oroszetal97-1}
{Orosz}, J., {Wade}, R.~A., \& {Harlow}, J.~J.~B. 1997, AJ, 114, 317

\bibitem[{{Orosz} {et~al.}(1999){Orosz}, {Wade}, {Harlow}, {Thorstensen},
  {Taylor}, \& {Eracleous}}]{oroszetal99-1}
{Orosz}, J.~A., {Wade}, R.~A., {Harlow}, J. J.~B., {et~al.} 1999, AJ, 117, 1598

\bibitem[{{Paczynski} \& {Sienkiewicz}(1983)}]{paczynski+sienkiewicz83-1}
{Paczynski}, B. \& {Sienkiewicz}, R. 1983, ApJ, 268, 825

\bibitem[{{Patterson}(1984)}]{patterson84-1}
{Patterson}, J. 1984, ApJS, 54, 443

\bibitem[{{Patterson}(1998)}]{patterson98-1}
---. 1998, PASP, 110, 1132

\bibitem[{{Perryman} {et~al.}(1998){Perryman}, {Brown}, {Lebreton}, {Gomez},
  {Turon}, {de Strobel}, {Mermilliod}, {Robichon}, {Kovalevsky}, \&
  {Crifo}}]{perrymanetal98-1}
{Perryman}, M.~A.~C., {Brown}, A.~G.~A., {Lebreton}, Y., {et~al.} 1998, A\&A,
  331, 81

\bibitem[{{Pinsonneault} {et~al.}(2002){Pinsonneault}, {Andronov}, \&
  {Sills}}]{pinsonneaultetal02-1}
{Pinsonneault}, M.~H., {Andronov}, N., \& {Sills}, A. 2002, in The Physics of
  Cataclysmic Variables and Related Objects, ed. B.~T. {G\"ansicke},
  K.~{Beuermann}, \& K.~{Reinsch} (ASP Conf. Ser. 261), 208--216

\bibitem[{{Politano}(1996)}]{politano96-1}
{Politano}, M. 1996, ApJ, 465, 338

\bibitem[{{Pollacco} \& {Bell}(1993)}]{pollacco+bell93-1}
{Pollacco}, D.~L. \& {Bell}, S.~A. 1993, MNRAS, 262, 377

\bibitem[{{Pollacco} \& {Bell}(1994)}]{pollacco+bell94-1}
---. 1994, MNRAS, 267, 452

\bibitem[{{Pottasch}(1996)}]{pottasch96-1}
{Pottasch}, S.~R. 1996, A\&A, 307, 561

\bibitem[{{Pounds} {et~al.}(1993){Pounds}, {Allan}, {Barber}, {Barstow},
  {Bertram}, {Branduardi-Raymont}, {Brebner}, {Buckley}, {Bromage}, {Cole},
  {Courtier}, {Cruise}, {Culhane}, {Denby}, {Donoghue}, {Dunford},
  {Georgantopoulos}, {Goodall}, {Gondhalekar}, {Gourlay}, {Harris}, {Hassall},
  {Hellier}, {Hodgkin}, {Jeffries}, {Kellett}, {Kent}, {Lieu}, {Lloyd},
  {McGale}, {Mason}, {Matthews}, {Mittaz}, {Page}, {Pankiewicz}, {Pike},
  {Ponman}, {Puchnarewicz}, {Pye}, {Quenby}, {Ricketts}, {Rosen}, {Sansom},
  {Sembay}, {Sidher}, {Sims}, {Stewart}, {Sumner}, {Vallance}, {Watson},
  {Warwick}, {Wells}, {Willingale}, {Willmore}, {Willoughby}, \&
  {Wonnacott}}]{poundsetal93-1}
{Pounds}, K.~A., {Allan}, D.~J., {Barber}, C., {et~al.} 1993, MNRAS, 260, 77

\bibitem[{{Prialnik} \& {Livio}(1985)}]{prialnik+livio85-1}
{Prialnik}, D. \& {Livio}, M. 1985, MNRAS, 216, 37

\bibitem[{{Rappaport} {et~al.}(1983){Rappaport}, {Joss}, \&
  {Verbunt}}]{rappaportetal83-1}
{Rappaport}, S., {Joss}, P.~C., \& {Verbunt}, F. 1983, ApJ, 275, 713

\bibitem[{{Rauch}(2000)}]{rauch00-1}
{Rauch}, T. 2000, A\&A, 356, 665

\bibitem[{{Raymond} {et~al.}(2003){Raymond}, {Szkody}, {Hawley}, {Anderson},
  {Brinkmann}, {Covey}, {McGehee}, {Schneider}, {West}, \&
  {York}}]{raymondetal03-1}
{Raymond}, S.~N., {Szkody}, P., {Hawley}, S.~L., {et~al.} 2003, AJ, 125, 2621

\bibitem[{{Ringwald}(1996)}]{ringwald96-1}
{Ringwald}, F.~A. 1996, in Cataclysmic Variables and Related Objects, ed.
  A.~{Evans} \& J.~H. {Wood}, IAU Coll. No. 158 (Dordrecht: Kluwer), 89--92

\bibitem[{{Ritter}(1986)}]{ritter86-2}
{Ritter}, H. 1986, A\&A, 169, 139

\bibitem[{{Ritter} \& {Kolb}(1998)}]{ritter+kolb98-1}
{Ritter}, H. \& {Kolb}, U. 1998, A\&AS, 129, 83

\bibitem[{{Rodgers} \& {Eggen}(1974)}]{rodgers+eggen74-1}
{Rodgers}, A.~W. \& {Eggen}, O.~J. 1974, PASP, 86, 742

\bibitem[{{Saffer} {et~al.}(1993){Saffer}, {Wade}, {Liebert}, {Green}, {Sion},
  {Bechtold}, {Foss}, \& {Kidder}}]{safferetal93-1}
{Saffer}, R.~A., {Wade}, R.~A., {Liebert}, J., {et~al.} 1993, AJ, 105, 1945

\bibitem[{{Schenker} \& {King}(2002)}]{schenker+king02-1}
{Schenker}, K. \& {King}, A.~R. 2002, in The Physics of Cataclysmic Variables
  and Related Objects, ed. B.~T. {G\"ansicke}, K.~{Beuermann}, \& K.~{Reinsch}
  (ASP Conf. Ser. 261), 242--251

\bibitem[{{Schmidt} {et~al.}(1995){Schmidt}, {Smith}, {Harvey}, \&
  {Grauer}}]{schmidtetal95-3}
{Schmidt}, G.~D., {Smith}, P.~S., {Harvey}, D.~A., \& {Grauer}, A.~D. 1995, AJ,
  110, 398

\bibitem[{{Schultz} {et~al.}(1993){Schultz}, {Zuckerman}, {Becklin}, \&
  {Barnbaum}}]{schultzetal93-1}
{Schultz}, G., {Zuckerman}, B., {Becklin}, E.~E., \& {Barnbaum}, C. 1993, BAAS,
  25, 824

\bibitem[{{Sills} {et~al.}(2000){Sills}, {Pinsonneault}, \&
  {Terndrup}}]{sillsetal00-1}
{Sills}, A., {Pinsonneault}, M.~H., \& {Terndrup}, D.~M. 2000, ApJ, 534, 335

\bibitem[{{Sing} {et~al.}(2001){Sing}, {Holberg}, {Howell}, {Barstow}, \&
  {Burleigh}}]{singetal01-1}
{Sing}, D.~K., {Holberg}, J.~B., {Howell}, S., {Barstow}, M.~A., \& {Burleigh},
  M. 2001, American Astronomical Society Meeting, 199, 0

\bibitem[{{Somers} {et~al.}(1996){Somers}, {Lockley}, {Naylor}, \&
  {Wood}}]{somersetal96-2}
{Somers}, M.~W., {Lockley}, J.~J., {Naylor}, T., \& {Wood}, J.~H. 1996, MNRAS,
  280, 1277

\bibitem[{{Spruit} \& {Ritter}(1983)}]{spruit+ritter83-1}
{Spruit}, H.~C. \& {Ritter}, H. 1983, A\&A, 124, 267

\bibitem[{{Spruit} \& {Taam}(2001)}]{spruit+taam01-1}
{Spruit}, H.~C. \& {Taam}, R.~E. 2001, ApJ, 548, 900

\bibitem[{{Stauffer}(1987)}]{stauffer87-1}
{Stauffer}, J.~R. 1987, AJ, 94, 996

\bibitem[{{Stephenson}(1960)}]{stephenson60-1}
{Stephenson}, C.~B. 1960, PASP, 72, 387

\bibitem[{{Stephenson} \& {Sanduleak}(1971)}]{stephenson+sanduleak71-1}
{Stephenson}, C.~B. \& {Sanduleak}, N. 1971, {Luminous stars in the Southern
  Milky Way}, Vol.~1 (Publications of the Warner \& Swasey Observatory)

\bibitem[{{Thorstensen} {et~al.}(1994){Thorstensen}, {Vennes}, \&
  {Shambrook}}]{thorstensenetal94-1}
{Thorstensen}, J.~R., {Vennes}, S., \& {Shambrook}, A. 1994, AJ, 108, 1924

\bibitem[{{Townsley} \& {Bildsten}(2002)}]{townsley+bildsten02-1}
{Townsley}, D.~M. \& {Bildsten}, L. 2002, in The Physics of Cataclysmic
  Variables and Related Objects, ed. B.~T. {G\"ansicke}, K.~{Beuermann}, \&
  K.~{Reinsch} (ASP Conf. Ser. 261), 31--40

\bibitem[{{Tweedy} {et~al.}(1993){Tweedy}, {Holberg}, {Barstow}, {Bergeron},
  {Grauer}, {Liebert}, \& {Fleming}}]{tweedyetal93-1}
{Tweedy}, R.~W., {Holberg}, J.~B., {Barstow}, M.~A., {et~al.} 1993, AJ, 105,
  1938

\bibitem[{{Tytler} \& {Rubenstein}(1989)}]{tytler+rubenstein89-1}
{Tytler}, D. \& {Rubenstein}, E. 1989, in White dwarfs, ed. G.~{Wegner}, IAU
  Coll. No. 158 (Heidelberg: Springer), 524

\bibitem[{{Vennes} \& {Thorstensen}(1994)}]{vennes+thorstensen94-1}
{Vennes}, S. \& {Thorstensen}, J.~R. 1994, ApJ Lett., 433, L29

\bibitem[{{Vennes} \& {Thorstensen}(1996)}]{vennes+thorstensen96-1}
---. 1996, AJ, 112, 284

\bibitem[{{Vennes} {et~al.}(1999){Vennes}, {Thorstensen}, \&
  {Polomski}}]{vennesetal99-2}
{Vennes}, S., {Thorstensen}, J.~R., \& {Polomski}, E.~F. 1999, ApJ, 523, 386

\bibitem[{{Verbunt} \& {Zwaan}(1981)}]{verbunt+zwaan81-1}
{Verbunt}, F. \& {Zwaan}, C. 1981, A\&A, 100, L7

\bibitem[{{Walsh} \& {Walton}(1996)}]{walsh+walton96-1}
{Walsh}, J.~R. \& {Walton}, N.~A. 1996, A\&A, 315, 253

\bibitem[{{Walton} {et~al.}(1993){Walton}, {Walsh}, \&
  {Pottasch}}]{waltonetal93-1}
{Walton}, N.~A., {Walsh}, J.~R., \& {Pottasch}, S.~R. 1993, A\&A, 275, 256

\bibitem[{{Warner}(1995)}]{warner95-1}
{Warner}, B. 1995, Cataclysmic Variable Stars (Cambridge: Cambridge University
  Press)

\bibitem[{{Wegner}(1973)}]{wegner73-1}
{Wegner}, G. 1973, MNRAS, 163, 381

\bibitem[{{Weidemann}(2000)}]{weidemann00-1}
{Weidemann}, V. 2000, A\&A, 363, 647

\bibitem[{{Wills} \& {Wills}(1974)}]{wills+wills74-1}
{Wills}, D. \& {Wills}, B.~J. 1974, MNRAS, 167, 79P

\bibitem[{{Wilson}(1953)}]{wilson53-1}
{Wilson}, R.~E. 1953, {General Catalogue of Stellar Radial Velocities}
  (Carnegie Institute Washington D.C.~Publication)

\bibitem[{{Wood} \& {Marsh}(1991)}]{wood+marsh91-1}
{Wood}, J.~H. \& {Marsh}, T.~R. 1991, ApJ, 381, 551

\bibitem[{{Wood} {et~al.}(1995){Wood}, {Robinson}, \& {Zhang}}]{woodetal95-3}
{Wood}, J.~H., {Robinson}, E.~L., \& {Zhang}, E.-H. 1995, MNRAS, 277, 87

\bibitem[{{Wood} \& {Saffer}(1999)}]{wood+saffer99-1}
{Wood}, J.~H. \& {Saffer}, R. 1999, MNRAS, 305, 820

\bibitem[{{Wood}(1995)}]{wood95-1}
{Wood}, M.~A. 1995, in White Dwarfs, ed. D.~{Koester} \& K.~{Werner}, LNP No.
  443 (Heidelberg: Springer), 41--45

\bibitem[{{Zuckerman} \& {Becklin}(1992)}]{zuckerman+becklin92-1}
{Zuckerman}, B. \& {Becklin}, E.~E. 1992, ApJ, 386, 260

\end{thebibliography}
\end{document}